\documentclass[prx,longbibliography,twocolumn,reprint,superscriptaddress,amsmath,floatfix]{revtex4-1}
\usepackage{graphicx}
\usepackage{amssymb}
\usepackage{multirow}
\usepackage[percent]{overpic}
\usepackage[svgnames]{xcolor}
\begin{document}

\title{Sr$_2$RuO$_4$, like doped cuprates and barium bismuthate, is a negative charge-transfer gap 
even parity superconductor with $\frac{3}{4}$-filled oxygen band}
\author{Sumit Mazumdar}
\affiliation{Department of Physics, University of Arizona
Tucson, AZ 85721}
\affiliation{Department of Chemistry and Biochemistry, University of Arizona, Tucson, AZ 85721}
\affiliation{College of Optical Sciences, University of Arizona, Tucson, AZ 85721}
\date{\today}
\vskip 1pc
\begin{abstract}
A comprehensive theory of superconductivity in Sr$_2$RuO$_4$ must simultaneously explain experiments 
that suggest even parity superconducting order and yet others that have suggested broken
time reversal symmetry. Completeness further requires that the theory is applicable to isoelectronic
Ca$_2$RuO$_4$, a Mott-Hubbard semiconductor that exhibits an unprecedented insulator-to-metal transition 
which can be driven by very small electric field or current, and also by doping with very
small concentration of electrons, leading to a metallic state proximate to ferromagnetism. A valence
transition model, previously proposed for superconducting cuprates [Phys. Rev. B {\bf 98}, 205153] is
here extended to Sr$_2$RuO$_4$ and Ca$_2$RuO$_4$. The insulator-to-metal transition is distinct from that
expected from the simple “melting” of the Mott-Hubbard semiconductor. Rather, the Ru-ions occur
as low-spin Ru$^{4+}$ in the semiconductor, and as high-spin Ru$^{3+}$ in the metal, the driving force behind
the valence transition being the strong spin-charge coupling and consequent large ionizaton energy
in the low charge state. Metallic and superconducting ruthenates are thus two-component systems
in which the half-filled high-spin Ru$^{3+}$ ions determine the magnetic behavior but not transport,
while the charge carriers are entirely on on the layer oxygen ions, which have an average charge
-1.5. Spin-singlet superconductivity in Sr$_2$RuO$_4$ evolves from the correlated lattice-frustrated 
3/4-filled band of layer oxygen ions alone, in agreement with quantum many-body calculations that have
demonstrated enhancement by electron-electron interactions of superconducting pair-pair correla-
tions uniquely at or very close to this filling [Phys. Rev. B {\bf 93}, 165110 and {\bf 93}, 205111]. Several
model-specific experimental predictions are made, including that spin susceptibility due to Ru-ions
will remain unchanged as Sr$_2$RuO$_4$ is taken through superconducting Tc.
\end{abstract}
\maketitle
\section{\bf Introduction.}
Sr$_2$RuO$_4$ has long been thought of as a chiral spin-triplet superconductor, with orbital parity p$_x$ $\pm$ $i$p$_y$ 
\cite{MacKenzie03a}.
This viewpoint has recently been challenged by multiple experiments
\cite{Mackenzie17a,Hicks14a,Taniguchi15a,Steppke17a,Hassinger17a,Luo19a,Pustogow19a} that are beginning to lead to a thorough 
re-examination of earlier experiments or their interpretations and theories. A few investigators have even suggested that the superconducting pairing might have 
even parity,
likely $d$-wave \cite{Hassinger17a,Sharma20a}. In the present theoretical work, which is an extension of a valence trasition model \cite{Mazumdar18a} recently postulated for
superconducting cuprates and doped barium bismuthate, (Ba,K)BiO$_3$, I posit that the peculiarities observed in Sr$_2$RuO$_4$ and the isoelectronic Ca$_2$RuO$_4$
should not be considered
in isolation, but that the unconventional behaviors of all these superconducting perovskite oxides, along with the pseudogap-like features 
\cite{Torre15a,Kim16a,Battisti17a} 
observed in electron-doped Sr$_2$IrO$_4$, can be understood within a common theoretical model. The key theoretical features of this 
theory are valence transition and negative charge-transfer gap, which in all cases are driven by a unique property
common to the materials of interest:
the essential cation in each material has a strong tendency to have true ionicity lower than the formal ionicity by a full integer unit. This is a strong correlations
effect that is missed in band or first principles calculations, which invariably find a single phase with mixed valence as opposed to distinct phases with 
nearly integer valences. 

It is useful to point out here that the concept of valence transition in systems consisting of donor and acceptor components
is widely accepted in the context of pressure, temperature and light-induced neutral-to-ionic transition in quasi-one-dimensional mixed-stack organic charge-transfer
solids \cite{Torrance81a,Kosihara90a,Masino17a}. The concept of quantum critical valence fluctuation in heavy fermions \cite{Scheerer18a} 
is qualitatively similar. In the superconducting oxides the consequence of valence transition is   
an insulator-to-metal transition (IMT) where the metallic state is different from  
that arising from simple ``doping'' or ``self-doping'' of the semiconductor,
as has been assumed until now. Within the proposed model the insulating phase has the usual M$^{n+}$(O$^{2-})_2$ intralayer unit cell 
composition, but the 
true (as opposed to formal) ionic composition in the pseudogap and metallic states is M$^{(n-1)+}$(O$^{1.5-})_2$. 
The pseudogap state (wherever applicable) and the ``normal'' 
state from which superconductivity (SC) emerges thus consist of a strongly correlated oxygen (O)-band in which nearly half the oxygens 
(as opposed to a few) have ionicities O$^{1-}$. The cations in their true ionicities are either closed-shell or exactly half-filled
and play no role in SC. 

The negative charge-transfer gap model is arrived at from heuristic arguments as opposed to direct computations, as the sheer number of many-body interactions and 
parameters that would enter such computations are enormously large. The relative magnitudes of the different parameters 
that would enter a complete theoretical model are known, however, which allows making the physical arguments for the transition. 
This approach is counter to existing theoretical approaches to Sr$_2$RuO$_4$, but as in the previous work \cite{Mazumdar18a}
it will be shown that (nearly) all the experimental observations that are difficult to explain within any other single theory can be explained relatively
easily once the
fundamental premise is accepted. 
The validity of the model can therefore be tested only by comparing theoretically
arrived at conclusions against existing experiments as well as new experiments that are performed to test predictions of the theory. It then becomes necessary to
list the full set of experiments that any theory that claims to be comprehensive should be able to explain at least qualitatively. This is the
approach that was taken in the earlier work on the cuprates \cite{Mazumdar18a} and is taken here for Sr$_2$RuO$_4$ and Ca$_2$RuO$_4$. 

In the next section I list what I believe are
the most challenging observations, including apparent contradictions, in (i) Sr$_2$RuO$_4$, and (ii) the isoelectronic Mott-Hubbard semiconductor Ca$_2$RuO$_4$,
which exhibits dramatic IMT induced by current \cite{Nakamura13a,Okazaki13b} and electron-doping \cite{Cao00a}, as well as exotic behavior in the
nanocrystal form \cite{Nobukane20a}. Existing theories of Sr$_2$RuO$_4$ largely attempt to determine the superconducting symmetry of the material while ignoring
the highly unusual behavior of Ca$_2$RuO$_4$. The approach here is to treat both systems on an equal footing.  
Following this I briefly present in section III the theory of what I term as
type I negative charge-transfer gap, as observed in doped cuprates, BaBiO$_3$ and (Ba,K)BiO$_3$, and doped Sr$_2$IrO$_4$ in octahedral environment. 
Although much of this has already been
presented in the earlier work \cite{Mazumdar18a}, it is necessary to repeat this briefly here to point out the unique common feature shared by Cu$^{1+}$, Bi$^{3+}$, 
Ir$^{3+}$ in octahedral environment, and Ru$^{3+}$. It is this shared feature that is the driver of an unusual IMT in these perovskite oxides. Section IV discusses the physical 
mechanism behind the valence transition that drives what I term as the type II negative charge-transfer gap in Sr$_2$RuO$_4$. Section V shows how all the experiments
listed in section II, in particular, (i) spin-singlet even parity SC and the confusing experimental observations on time reversal symmetry breaking
in Sr$_2$RuO$_4$, and (ii) the current and
electron-doping induced metallicicity in Ca$_2$RuO$_4$ and perhaps even the SC claimed to have been observed in this system 
{\it can be simultaneously understood within the theoretical model}. Finally in section VI I make experimentally testable predictions on Sr$_2$RuO$_4$ that are completely specific to the negative charge-transfer gap model. 
Section VII presents the conclusions, focusing in particular on the unique features of the correlated $\frac{3}{4}$-filled band. 

\section{Theoretical Challenges}
\label{challenges}

\subsection {Experimental puzzles, Sr$_2$RuO$_4$.}
\label{SRO}

\noindent {\it (i) T$_c$ enhancement under uniaxial pressure.} The superconducting critical temperature T$_c$ in Sr$_2$RuO$_4$ crystals
is strongly enhanced upon the application of uniaxial pressure along the [100] direction \cite{Hicks14a,Taniguchi15a,Steppke17a}, even as
hydrostatic pressure suppresses T$_c$. Starting from the ambient pressure value of 1.5 K, T$_c$ reaches a peak value of 3.4 K at uniaxial compression of
0.6\%, following which T$_c$ decreases again. Based on band structure calculations it has been suggested the peak in T$_c$ corresponds
to the compression at which the Fermi level crosses the van Hove singularity \cite{Steppke17a}. Theoretically predicted splitting of the transition temperatures due to separate
p$_x$ and p$_y$ components were not observed. The superconducting
transition at the maximum T$_c$ is very sharp, allowing precise determinations of the upper critical fields
for magnetic fields both along the intra-plane [100] direction (H$_{c2||a}$) as well as perpendicular to the plane  (H$_{c2||c}$).
While H$_{c2||c}$ is enhanced by more than a factor of 20 relative to unstrained Sr$_2$RuO$_4$, in-plane H$_{c2||a}$  is enhanced by only a factor of 3. Importantly,
for the spins lying in the (2D) plane, neither orbital limiting nor Pauli limiting should apply to H$||a$, and H$_{c2||a}$/H$_{c2||c}$ should be infinite
within the existing spin and orbital characterization of the superconducting state.
{\it The observed ratio of only about 3 in the strained material casts severe doubts about the chiral $p$-wave symmetry.} 

In view of what follows in Section IV, it is pointed out here that strong pressure-induced enhancement of T$_c$ is also found 
in Ce-based heavy fermion superconductors, and is ascribed to critical valence fluctuations by some investigators \cite{Scheerer18a}.

\noindent {\it (ii) $^{17}$O NMR.} The earlier experiment that had given the most convincing evidence for triplet SC was based on the
measurement of the O-ion Knight-shift as a function of temperature \cite{Ishida98a}. No change in spin susceptibility was detected
as the sample was taken through the critical temperature T$_c$. Luo {\it et al.} \cite{Luo19a} and Pustogow {\it et al.} \cite{Pustogow19a} have repeated the
$^{17}$O NMR measurements in uniaxially compressed Sr$_2$RuO$_4$ for
different strain levels \cite{Hicks14a,Taniguchi15a,Steppke17a}, inclusive of the complete range of T$_c=1.5-3.4$ K. Reduction in the Knight
shift, and therefore drop in the spin susceptibility have been found for all strains, including for the unstrained sample \cite{Pustogow19a}. Most importantly,
{\it the NMR study finds no evidence for a transition between different symmetries.} The experiment conclusively precludes
p$_x$ $\pm$ $i$p$_y$ triplet pairing, and leaves open the possibilities of helical triplet pairings, spin-singlet d$_{xy}$ or
d$_{x^2-y^2}$ pairings and chiral $d$-wave pairing. 

\noindent {\it (iii) Breaking of time reversal symmetry?} Muon-spin rotation \cite{Luke98a} and magneto-optic polar Kerr rotation \cite{Xia06a} measurements
had suggested that time-reversal symmetry is broken upon entering the superconducting state. This conclusion has been contradicted by the observation
that the Josephson critical current is invariant under the inversion of current and magnetic fields \cite{Kashiwaya19a}. It is relevant in this context that
the polar Kerr effect is also seen in hole-doped cuprates inside the pseodogap, and while originally this was also ascribed to time reversal symmetry breaking,
it has been later ascribed to two-dimensional (2D) chirality \cite{Karapetyan12a}. Recent muon relaxation experiments on uniaxially stressed Sr$_2$RuO$_4$
have found that the onset temperatures of time reversal symmetry breaking T$_{TRSB}$ and superconducting T$_c$ are different \cite{Grinenko20a}. 
The authors also found magnetic order in Sr$_2$RuO$_4$ under high stress, but ignore this, and ascribe the difference between T$_{TRSB}$ and T$_c$ to
two-component SC. Within the existing Ru-centric theories of SC this is the only possibility. Within the negative charge-tranfer gap model 
T$_{TRSB}$ is associated with magnetism involving the Ru-ions and T$_c$ to the O-ions (see Section V). 
 
\noindent {\it (iv) Magnetocaloric and thermal conductivity measurements.} Magnetocaloric measurements have found that the superconductor-to-metal transition
in the unstrained material at T $\simeq 0.5$T$_c$ is first order, indicating that the pair-breaking is Pauli-limited, {\it i.e.}, pairing is spin-singlet \cite{Yonezawa13a}. 
Among the symmetries not precluded by the $^{17}$O NMR experiment \cite{Pustogow19a} the helical triplet orders and the chiral $d$-wave order have horizontal
nodes while the $d_{xy}$ and $d_{x^2-y^2}$ orders have vertical nodes. Recent thermal conductivity measurements have found evidence for vertical line nodes
consistent with  $d$-wave pairing \cite{Hassinger17a}.

%
%
\subsection{Experimental puzzles, Ca$_2$RuO$_4$}
\label{CaRO}

Replacement of Sr with Ca generates the isoelectronic Ca$_2$RuO$_4$ as well as the ``doped'' compounds Ca$_{2-x}$Sr$_x$RuO$_4$, $0\leq x\leq 2$.
Ca$_2$RuO$_4$ is an antiferromagnetic
semiconductor with energy gap between 0.2 - 0.4 eV at low temperatures \cite{Braden98a,Friedt01a,Fukazawa01a,Nakatsuji04a}. Ru-ions in this compound
have ionicity +4 and are in the low-spin state, with 4$d$ orbital occupancy $t_{2g}^4$. N\'eel temperature of
113 K and paramagnetic semiconductor to paramagnetic metal transition at $\sim 360$ K show that the system is a Mott-Hubbard semiconductor.
The mechanism of the Mott-Hubbard IMT remains controversial but one popular
mechanism involves increased $d_{xy}$-occupancy to upto 2 electrons due to Jahn-Teller distortion, with Hund's rule coupling leading
to 1 electron each in the $d_{xz}$ and $d_{yz}$ orbitals \cite{Fukazawa01a,Liebsch07a,Gorelov10a}. The half-filled nature of the $d_{xz}$ and $d_{yz}$ bands then lead to
Mott-Hubbard semiconducting behavior. The Mott-Hubbard transition is accompanied with structural distortions involving tilts and rotations of the
RuO$_6$ octahedra, with the layer (apical) Ru-O bonds becoming shorter (longer) in the metallic phase (the semiconducting antiferromagnetic and the metallic
phases are commonly labeled as $S$ and $L$, respectively) \cite{Friedt01a}.
Similar structural changes are also seen in pressure-induced IMT transition, where the metallic state is
found to be proximate to ferromagnetism \cite{Nakamura02b}.
``Doping''-induced IMT occurs in Ca$_{2-x}$Sr$_x$RuO$_4$ for $x>0.2$. Importantly,
the high temperature metallic phase in Ca$_2$RuO$_4$, the paramagnetic metallic phase $0.5 \leq x \leq 1.5$ and the $x=2$ phase pure Sr$_2$CuO$_4$ are all
structurally different. Latest experiments have found completely different IMT that has not been seen so far in
any other Mott-Hubbard semiconductor, as described below.

\noindent{\it (i) Current-induced IMT.} An unprecedented electric field-induced IMT, with a lower threshold field of 40 V/cm, tiny compared to the known 
semiconducting gap of 0.2 - 0.4 eV, has been found in Ca$_2$RuO$_4$ \cite{Nakamura13a,Okazaki13b}. 
Strong current-induced diamagnetism, the origin of which is not understood, appears in the semimetallic state reached with a current density as low as 1 A/cm$^2$ \cite{Sow17a}.
The IMT is not due to Joule heating, 
as evidenced by Raman spectroscopy \cite{Fursich19a} and the nucleation of the metallic phase at the negative electrode \cite{Zhang19a}.   
The crystal structure of the current-induced semimetallic state is different from the equilibrium 
state reached by applications of temperature, pressure or strain \cite{Bertinshaw19a}.
The rapid suppression of the antiferromagnetic order and resistance even at the smallest currents
is accompanied by strong lattice distortions \cite{Zhang19a,Bertinshaw19a,Zhao19a}
(reference \onlinecite{Zhao19a} however does not find diamagnetism). The very strong ``coupling'' between the current and lattice structure has led to the suggestion
that the $t_{2g}$ orbital occupancies of Ru$^{4+}$ ions in the semiconductor and the current-carrying state are different \cite{Zhao19a}. 

\noindent {\it (ii) IMT induced by La substitution.} There exist remarkable yet hitherto unnoticed similarities between IMT induced by current and electron doping
by substitution of La for Ca. The ionic radii of Ca$^{2+}$ (1.00 \AA) and La$^{3+}$ (1.03 \AA) are nearly equal. The consequences of La-substitition are 
more dramatic than Sr-substitution, even though Sr$_2$RuO$_4$ is isoelectronic with Ca$_2$RuO$_4$ \cite{Cao00a}.
For $x$ as small as 0.005 in Ca$_{2-x}$La$_x$RuO$_4$, the temperature at which the IMT occurs drops by nearly 80 K and resistance decreases by 
two orders of magnitude throughout the measured
temperature range. The system is also ferromagnetic below 125 K. For $x=0.1$, the system is metallic down to 2 K and the resistance drops to
10$^{-4}$ ohm-cm from $\sim 10^9$ ohm-cm at $x=0$. The IMT is accompanied by large increases in Pauli paramagnetic susceptibility and the 
coefficient $\gamma$ of electronic
specific heat ($\gamma$ increases from 16 mJ/mole K$^2$ to 90 mJ/mole K$^2$ from $x=0.01$ to $x=0.035$) as well as large structural distortions. 
The unusually large increase in conductivity with the slightest of electron-doping, or under very small electric fields both indicate 
that {\it the ground state of the undoped compound is very close to an exotic instability}. 

\noindent{\it (iii) Coexisting SC and ferromagnetism?} A very recent work on nanofilm (as opposed to bulk) crystals of Ca$_2$RuO$_4$ has reported an even more 
perplexing phenomenon, possible co-appearance of ferromagnetism and superconducting order \cite{Nobukane20a}. The nanofilm crystals had the same $L$ phase structure as
metallic Ca$_2$RuO$_4$ as opposed to the $S$ structure in the bulk antiferromagnetic phase. 
SC seems to appear within a ferromagnetic phase whose T$_{Curie} \sim 180$ K. The authors find a diamagnetic component within the ferromagnetic phase
above the superconducting T$_c$ that is larger than that found in reference 
\onlinecite{Sow17a}, and ascribe this to preformed Cooper pairs \cite{Nobukane20a}. The diamagnetism is enhanced by ferromagnetism in some of the samples, which 
led to the conclusion of coexistence of SC and ferromagnetism. The authors ascribe the observations to chiral $p$-wave SC, which would however contradict the
$^{17}$O-NMR experiment in Sr$_2$RuO$_4$ \cite{Pustogow19a}. 
I point out in Section III that these experiments are reminescent of the appearance of SC in undoped thin film 
T$^\prime$ cuprates \cite{Naito16a}.      


\subsection{Summary}
\label{summary}
To summarize, any theory of Sr$_2$RuO$_4$ must simultaneously explain the apparent contradictions between the $^{17}$O NMR experiments \cite{Luo19a,Pustogow19a} 
and the muon experiments \cite{Luke98a,Grinenko20a},
and also between experiments that find preservation \cite{Kashiwaya19a} as well as breaking \cite{Luke98a,Grinenko20a} of time reversal symmetry. 
Proper understanding of the 
current \cite{Nakamura13a,Okazaki13b,Sow17a,Zhang19a,Bertinshaw19a,Zhao19a} and electron-doping \cite{Cao00a} induced IMTs in Ca$_2$RuO$_4$, 
{\it including of the similarities in the observations in the two cases}, are essential, 
as the detailed mechanisms of these transitions likely reveal the origin of the
difference between Ca$_2$RuO$_4$ and Sr$_2$RuO$_4$. Band calculations ascribe the difference to lattice distortions; however, the distortions themselves
can be {\it consequences} of different Madelung stabilization energy contributions to the total energies of Ca$_2$RuO$_4$ and Sr$_2$RuO$_4$ (see section V), 
given the large differences in ionic radii of  Ca$^{2+}$ and Sr$^{2+}$. The proximity
of ferromagnetism to the superconducting state (or even their coexistence in nanofilm Ca$_2$RuO$_4$ \cite{Nobukane20a}) is suggested from multiple experiments. A consistent 
explanation of all of the above features within a single theory is currently lacking.

\section{\bf Cuprates, bismuthate and iridates}
\label{otherSC}
The central postulate of the present work is that the peculiarities observed in Sr$_2$RuO$_4$
should not be considered in isolation, as equally
perplexing mysteries persist with the other perovskite superconductors cuprates and (Ba,K)BiO$_3$, and in the pseudogap-like state \cite{Torre15a,Kim16a,Battisti17a} 
in electron-doped Sr$_2$IrO$_4$. I argue in this subsection
that the failure to arrive at a comprehensive
theory in every case stems from focusing on {\it cation-centric Hamiltonians} (for e.g., the single-band Hubbard model for cuprates).
In the following I first list the experiments in the cuprates, (Ba,K)BiO$_3$ and Sr$_2$IrO$_4$ that most strongly argue against
cation-centric Hamiltonians. Following this a brief presentation of the theory behind negative charge-transfer gap \cite{Mazumdar18a} in these systems
is presented.

\subsection{The need to go beyond cation-centric models, experiments}

(i) The simultaneous breaking of rotational and translational symmetries in the hole-doped T-phase cuprates, accompanied by
intra-unit cell O-ion inequivalency \cite{Wu15a,Kohsaka12a,Comin15a,Reichardt18a} illustrate most strongly the need to incorporate the O-ions explicitly in any
starting theoretical description of the
cuprates. With the T$^\prime$ cuprates, the most peculiar features are (i) the very robust antiferromagnetism in
the ``usual'' electron-doped materials \cite{Armitage10a}, (ii) the
appearance of SC at {\it zero doping} nevertheless in specially prepared thin film samples, with T$_c$ higher than the maximum T$_c$ in the
``usual'' materials \cite{Naito16a} (see also discussion of possible SC in undoped nanocrystals of Ca$_2$RuO$_4$ in Section IIB), and (iii)
charge-order with {\it nearly the same periodicity as the hole-doped cuprates} \cite{SilvaNeto15a,SilvaNeto16a}. Taken together,
these observations present the following conundrum. On the one hand,
inequivalent O-ions in the hole-doped compounds in the charge-ordered state from which the superconducting states emerge require that the O-ions are
included in any attempt to construct a comprehensive theoretical model. On the other, the apparent symmetry between the hole- and electron-doped
compounds (in so far as SC is concerned) {\it requires} explanation within a single-band model, since electron-hole symmetry is absent within 
multiband models.

(ii) Negative charge-transfer gap in BaBiO$_3$ is already recognized. The semiconducting gap in undoped
BaBiO$_3$ within traditional cation-centric models \cite{Rice81a} had been ascribed to a charge-density wave (CDW) consisting of alternate Bi$^{3+}$ and Bi$^{5+}$-ions. SC in Ba$_{1-x}$K$_x$BiO$_3$
within these models emerges from doping the parent Bi-based CDW. Recent theoretical and experimental demonstrations \cite{Plumb16a,Khazraie18a}
of the occurrence of Bi-ions exclusively as Bi$^{3+}$
show convincingly that the existing theories of SC are simplistic. {\it There is also no explanation of the limitation of
SC \cite{Plumb16a,Pei90a} to K-concentration $0.37 \leq x \leq 0.5$ in Ba$_{1-x}$K$_x$BiO$_3$,} an issue to which we return later.

(iii) Sr$_2$IrO$_4$ has attracted strong attention in recent years as an effective square lattice Mott-Hubbard insulator. The active layer consists of IrO$_2$ unit cells with nominally tetravalent Ir$^{4+}$. The $d$-electron
occupancy is t$_{2g}^5$ as a consequence of large crystal field stabilization energy (CFSE). The $t_{2g}$ orbitals are split by 
spin-orbit coupling  into lower twofold degenerate total angular momentum J$_{eff}=\frac{3}{2}$ levels and an upper
nondegenerate narrow J$_{eff}=\frac{1}{2}$ level \cite{Kim08a}.
Occupancy of the latter by a single unpaired electron explains the Mott-Hubbard like behavior of
undoped Sr$_2$IrO$_4$. 
Remarkable similarities \cite{Yan15a,Torre15a,Kim16a,Battisti17a} are found between hole-doped cuprates
in the pseudogap phase and and electron-doped Sr$_2$IrO$_4$. Following the vanishing of the Mott-Hubbard gap at $\sim 5$\% doping there appears a $d$-wave like gap
in the nodal region, with strong deviation in the antinodal region, where the gap is much larger \cite{Battisti17a}, exactly as in the cuprates \cite{Hashimoto14a}.
Theoretical attempts to explain these observations borrow heavily from the single-band Hubbard model description for cuprates,
which, we have pointed out, is at best incomplete.

\subsection{The need to go beyond cations-centric models, theory} 

It is useful to point out a crucial recent theoretical development. Convincing proof of the 
absence of SC in the weakly doped 2D Hubbard Hamiltonian with nearest neighbor-only hoppings has been found from a comprehensive study that used two different complementary many-body
techniques \cite{Qin19a}. While reference \onlinecite{Qin19a} leaves open the
possibility that SC might still appear within a more complex Hubbard model with next-nearest neighbor hopping, calculations 
of the Hubbard $U$-dependence of superconducting pair-pair correlations in triangular lattices
preclude SC in the carrier concentration range 0.75 - 0.9 \cite{Gomes16a}, which has been thought to be appropriate for cuprates.

\subsection{Type I negative charge-transfer gap: Cations with closed shells}
\label{negative-gap}
The traditional approach to arriving at phenomenological minimal Hamiltonians for complex oxides
{\it assumes} that the nominal and true charges (ionicities) of the active cation and the O-ions are the same. The known example of BaBiO$_3$ (see above) already indicates 
that this can be incorrect. Physical understanding of the distinction between true versus formal charge is best obtained within strongly correlated ionic models 
\cite{Mazumdar78a,Nagaosa86a,Pati01a,Masino17a}. What follows merely requires small 
electron or hole hoppings between the central cation (Bi in BaBiO$_3$) and the O-anion, relative to the largest many-electron interactions. For any cation M that can exist in two different valence states M$^{n+}$ and M$^{(n-1)+}$, the true ionic charge of the
oxide is determined by the inequality \cite{Mazumdar18a}
\begin{equation}
I_n + A_2 + \Delta E_{M,n} + \Delta(W) \gtrless 0
\label{ionicity}
\end{equation}
where $I_n$ is energy of the $n$th ionization (M$^{(n-1)} \to$ M$^{n+}$ + $e$) and A$_2$ is the second electron-affinity of O.
Here $\Delta E_{M,n} = E_{M,n}-E_{M,n-1}$, where $E_{M,n}$ is
the Madelung energy of the solid with the cation charge of $+n$.
$\Delta(W)$ = $W_n - W_{n-1}$, where $W_n$ and $W_{n-1}$ are the gains in total one-electron delocalization (band) energies of states with cationic charges $+n$ and $+(n-1)$, respectively.
$I_n$ and $A_2$ are positive,  while $\Delta E_{M,n}$ is negative.
Note that, (i) $W_n$ and $W_{n-1}$ are both negative, (ii) for cation charge of $+n$ there are very few charge carriers, while for cation charge $+(n-1)$ a large fraction 
of the O-ions (nearly half) are O$^{1-}$ and the number of charge carriers is far larger, making $\Delta(W)$ positive. 
The two largest quantities in Eq.~\ref{ionicity}, $I_n$ and $\Delta E_{M,n}$, have
opposite signs and magnitudes several tens of eV or even larger (see below), and are many times larger than A$_2$ and $\Delta(W)$, which are both a few eV. 
This introduces the possibility of distinct quantum states with nearly integer valences, as opposed to mixed valence \cite{Masino17a,Mazumdar18a}.
For a smaller left hand side in Eq.~\ref{ionicity} the ground state occurs as predominantly M$^{n+}$; for a larger
left hand side M$^{(n-1)+}$ dominates the ground state. Distinct ground states are outside the scope of band theory, where the emphasis is only on $W_n$.
%
%
%

{\it Although in the above 
competition is assumed between M$^{(n-1)+}$ and M$^{n+}$, Eq.~\ref{ionicity} applies equally, if not even more strognly, to the case where the 
competition is between M$^{(n-1)+}$ and M$^{(n+1)+}$, as is true for BaBiO$_3$, where the competition is between Bi$^{3+}$ and Bi$^{5+}$}.

Since $\Delta E_{M,n}$ is nearly independent of the detailed nature of M within a given row of the periodic
table, it follows that for {\it unusually large I$_n$} oxides will have strong tendency to be in the
ionic state M$^{(n-1)+}$. This conclusion immediately explains the occurrence of only Bi$^{3+}$ in BaBiO$_3$. As shown in Fig.~1,
Bi$^{3+}$ with closed shell configuration ([Xe]$4f^{14}5d^{10}6s^2$) has unusually large ionization energy among the
three consecutive $p$-block elements Pb, Bi and Po in the periodic table.
\begin{figure}
\includegraphics[width=3.0in]{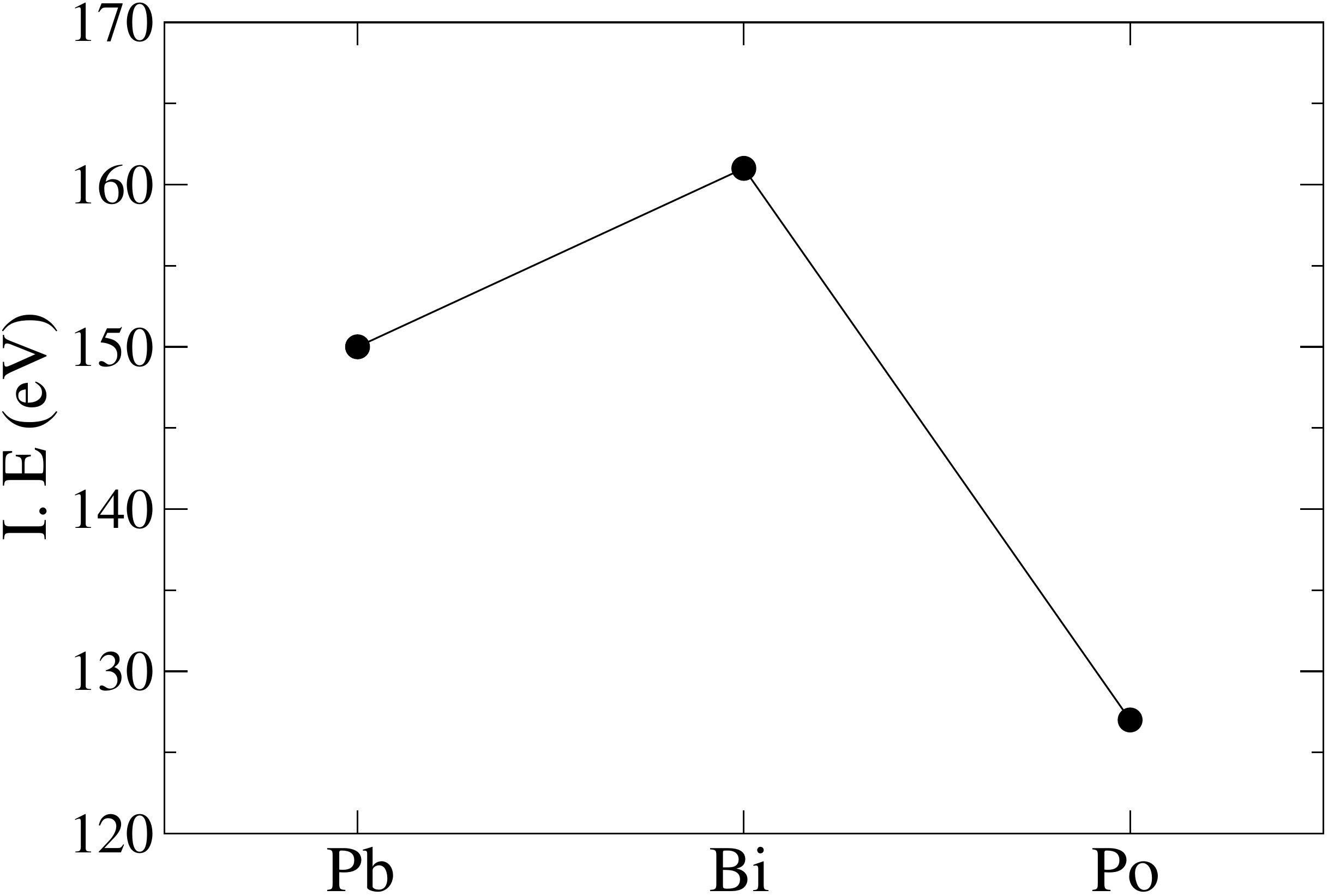}
\caption{Fourth ionization energies of Pb, Bi and Po (from www.webelements.com) The absence of Bi$^{5+}$ in BaBiO$_3$ is due to the exceptional stability of Bi$^{3+}$.
The plot of second ionization energies of 3$d$ transition metals shows a similar peak at Cu because of the closed-shell nature of Cu$^{1+}$,
see Fig.~1 in reference \onlinecite{Mazumdar18a} and Fig.~2(a).}
\label{IE}
\end{figure}

As with Bi$^{3+}$, unusually large $n$th ionization energy is a characteristic of all closed-shell M$^{(n-1)+}$. Additionally,
for systems already close to the boundary of the inequality \ref{ionicity}, external variables such as temperature, pressure or doping can change
$\Delta E_{M,n}$ or $\Delta(W)$ enough to lead to first order transition from
open-shell M$^{n+}$ to closed-shell M$^{(n-1)+}$. 
The most well-known among such valence transitions are the temperature, pressure and
light-induced neutral-to-ionic transitions in the
family of mixed-stack organic charge-transfer solids, which have been known for more than four decades \cite{Masino17a}.
Reference \onlinecite{Mazumdar18a} postulated that exactly such a
dopant-induced Cu$^{2+} \to$ Cu$^{1+}$ valence transition occurs at the pseudogap transition in the hole-doped
cuprates, and at the antiferromagnet-to-superconductor transition in the electron-doped cuprates.
The driving forces behind the transition are the unusually high second ionization energy
of closed-shell Cu$^{1+}$ with closed-shell electron configuration 3$d^{10}$ (see Fig.~1 in reference \onlinecite{Mazumdar18a}),
as well as contribution from $\Delta(W)$ in the doped state. The latter favors the lower
ionicity, because of the far greater number of charge carriers in this state.

The charge-transfer gap following the valence transition is the excitation Cu$^{1+}$O$^{1-} \to$ Cu$^{2+}$O$^{2-}$, opposite to
the excitation in the undoped semiconducting state, and is therefore ``negative''.
Following valence transition doped cuprates consist of an effective nearly $\frac{1}{4}$-filled
O hole-band ($\frac{3}{4}$-filled electron band) with the closed-shell Cu$^{1+}$ playing no significant role \cite{Mazumdar18a}.
The lattice of O-ions is frustrated (see Fig.~4). It has been shown from numerically accurate exact diagonalization,
quantum Monte Carlo and Path Integral Renormalization Group calculations, a density wave of Cooper pairs as well as a superconducting
state occur naturally and uniquely within the frustrated $\frac{1}{4}$-filled (and $\frac{3}{4}$-filled) band Hubbard Hamiltonian \cite{Gomes16a,DeSilva16a,Clay19a}. 
This effective {\it anion-centric
one-band model} can describe both hole- and
electron-doped cuprates, and aside from explaining correlated-electron SC, is able to give detailed physical understandings of the spatial broken symmetries in the
hole-doped cuprates, the unusual stability of the antiferromagnetic phase in the standard T$^\prime$ compounds as well as the appearance of SC in undoped thin film
T$^\prime$ cuprates \cite{Mazumdar18a}.

Closed-shell characters of Cu$^{1+}$ and Bi$^{3+}$ are true independent of crystal structure.  In octahedral complexes with large CFSE unusually large 
ionization energy will be true for cations with electron occupancy of $t_{2g}^6$. This tendency would be strongest with 
5$d$ cations with CFSE much larger than for 3$d$ and 4$d$ cations. Reference  \onlinecite{Mazumdar18a} therefore proposed that the transition to the 
pseudogap state in electron-doped Sr$_2$IrO$_4$ is a consequence the valence transition from Ir$^{4+}$ with open-shell configuration $t_{2g}^5$ to 
Ir$^{3+}$ with closed-shell $t_{2g}^6$. As in the cuprates this
would again imply a nearly $\frac{1}{4}$-filled oxygen hole-band, which would have the tendency to the same charge-ordering and hence the same $d$-wave like gap.
Strong support for this viewpoint comes from the {\it known} true charge distribution in octahedral IrTe$_2$. Even as the nominal charges  are Ir$^{4+}$(Te$^{2-})_2$,
the true ionic charges are accepted \cite{Oh13a,Fang13a} to be Ir$^{3+}$(Te$^{1.5-})_2$. 

\section{\bf Type II negative charge-transfer gap: cations with half-filled shells.}
\label{TypeII}

As seen in the previous section, negative charge-transfer is very likely with closed-shell cations and can be driven by both $\Delta E_{m,n}$ or $\Delta (W)$,
thereby introducing a new mechanism for IMT. In the following I discuss how a similar IMT can occur in complexes where formal charges
correspond to nearly half-filled $d$-shells.

\subsection{The unusally large ionization energies of half-filled ions.}
\label{half-filledIE}

Beyond completely closed-shell cations, isolated free ions that are exactly half-filled
also have unusually large ionization energies. For the $d$-block elements this is true for ions with electron configurations
$d^5$, which as free ions occur in their high-spin configurations because of Hund's coupling. Fig.~2(a) shows that the second ionization energy of
of Cr (Cr$^{1+} \to$ Cr$^{2+} + e$), the third ionization energy of Mn (Mn$^{2+} \to$ Mn$^{3+}$ + $e$) , and the 4th
ionization energy of Fe (Fe$^{3+} \to$ Fe$^{4+}+ e)$ are all significantly larger than those for similarly charged cations neighboring in the periodic table.
The behavior seen in Fig.~2(a) for free ions remains true for octahedral complexes of 3$d$ elements where CFSE is weak to moderate 
and Hund's coupling dominates. Thus octahedral complexes of
Mn$^{2+}$ are almost universally high-spin, while complexes of Fe$^{3+}$ are often
high-spin. This co-operative behavior emerges from the close coupling between high ionization energy and Hund's coupling in the 3$d$ series.
For the same ionic charge CFSE variation is $3d < 4d < 5d$. Thus the much larger $5d$ CFSE dominates over Hund's coupling in Ir,
and the closed-shell nature of octahedral Ir$^{3+}$ drives the Ir$^{4+}$ $\to$ Ir$^{3+}$ transition. 
Behavior intermediate between $3d$ (where Hund's coupling dominates) and $5d$ (where CFSE dominates) can emerge for $4d$, as discussed below.
   
\subsection {Valence transition and negative charge transfer gap in Sr$_2$RuO$_4$}

The nominal charge of the Ru-ion in Sr$_2$RuO$_4$ and Ca$_2$RuO$_4$ is Ru$^{4+}$ with four $d$-electrons. The ion is assumed to be in the low spin state 
in all prior theoretical work (two experimental studies have however claimed high-spin Ru$^{4+}$ in Sr$_2$RuO$_4$ \cite{Itoh95a} and metallic SrRuO$_3$ \cite{Grutter12a}).
Fig.~2(b) plots the fourth ionization energies of the 4$d$ free elements. As expected, the ionization energy of the isolated Ru$^{3+}$ ion is exceptionally large compared to
those of Tc$^{3+}$ and Rh$^{3+}$, with the differences (4.0 eV and $\sim 3$ eV, respectively) larger than the difference in the ionization energies of
Cu$^{1+}$ and Zn$^{1+}$ ($\sim 2$ eV). Should the true charge on the Ru-ions in Sr$_2$RuO$_4$ be +3 instead of the nominal +4, it would imply that Sr$_2$RuO$_4$
lies in the same class of materials as bismuthate, cuprate and Sr$_2$IrO$_4$.

\begin{figure}
\includegraphics[width=3.0in]{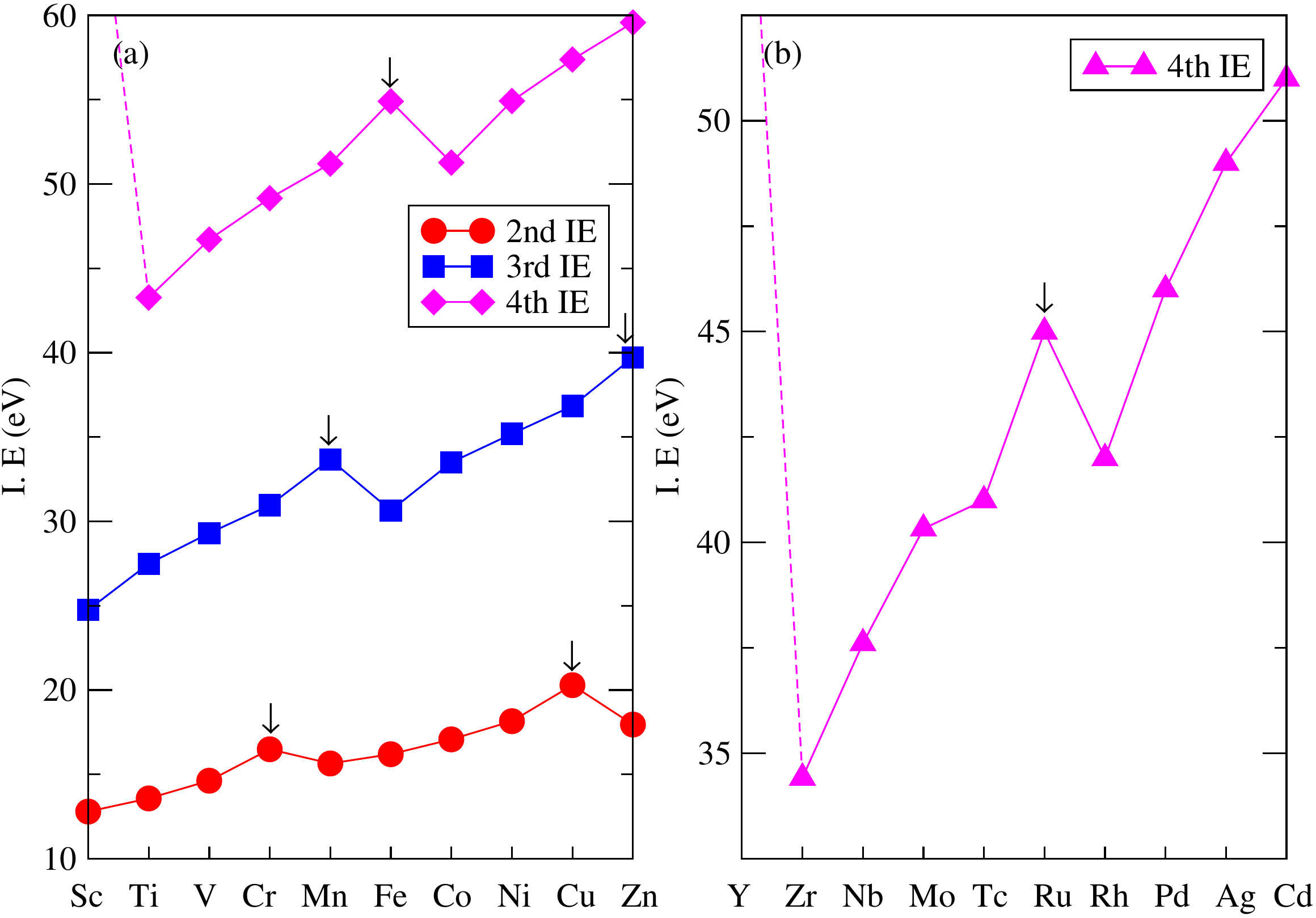}
\caption{(a) 2nd, 3rd and 4th ionization energies of 3$d$ transition metals. Note the local maxima on half-filled ions in each case. 
(b) 4th ionization energies of 4$d$ transition metals. Both sets of data are from www.webelements.com. 
The scales along the y-axes are different in (a) and (b), but the difference in the ionization energies of Ru$^{3+}$ and its
immediate neighbors in the periodic table is larger that that between Cu$^{1+}$ and Zn$^{1+}$.}
\label{1/2-filled}
\end{figure}

The charge-spin
coupling implied in Figs. 2(a) and (b) introduces a new mechanism for IMT in $d^4$-based systems, as presented below. As in the cuprates and bismuthates I compare the 
relative energies of the Sr$_2$RuO$_4$ crystal with Ru-ions in charge-state Ru$^{3+}$ (hereafter labeled as $|III\rangle$) versus 
charge-state Ru$^{4+}$ (labeled as $|IV\rangle$), with the additional condition that the Ru$^{3+}$-ions are assumed to be in the high-spin state (further
ionization of which is energetically costly as seen in Fig.~2(b)) while the    
Ru$^{4+}$-ions are low-spin.

Beyond the terms already included in Eq.~\ref{ionicity}, interactions that determine the relative
energies of $|III\rangle$ and $|IV\rangle$ include:
$U(CFSE:j)$, $U(Coulomb;j)$ and $U(exchange:j)$, $|j\rangle$ = $|III\rangle$ and $|IV\rangle$, where $U(CFSE:j)$ is the crystal field stabilization energy;
$U(Coulomb;j)$ the direct Coulomb repulsions between electrons occupying the $d$-orbitals (Hubbard repulsions between
electrons occupying the same as well as different $d$-orbitals); and $U(exchange:j)$ is the Hund's exchange energy.
The inequality that determines the competition between high-spin Ru$^{3+}$ and low-spin Ru$^{4+}$ is, 
\begin{eqnarray}
I_n + A_2 + \Delta E_{M,n} + \Delta(W) \\
\nonumber
+ \Delta_{CF}+\Delta_C+\Delta_J \gtrless 0 \\
\nonumber
\label{inequality2}
\end{eqnarray}
In the above $n=4$, 
and A$_2$, $\Delta E_{M,n}$ and  $\Delta(W)$ have the same meanings as in Eq.~\ref{ionicity}.
$\Delta_{CF}=U(CFSE:III)-U(CFSE:IV)$ is negative ($U(CFSE:III)\simeq0$) and favors state $|IV\rangle$ over state $|III\rangle$,
while $\Delta_J=U(exchange:III)-U(exchange:IV)$ is positive and favors state $|III\rangle$ over state $|IV\rangle$. It is difficult to estimate $\Delta_C=U(Coulomb;III)-U(Coulomb;IV)$; 
it is assumed to be small
relative to the larger $\Delta_{CF}$ and $\Delta_J$, the competition between which determine the relative stabilities of high versus low-spin in semiconductors. 
As with Eq.~\ref{ionicity}, a smaller (larger) left hand side favors Ru$^{4+}$ (Ru$^{3+}$).

The following are now pointed out:

(i) First-principles calculations for the cuprates have consistently determined large direct O-O hoppings $t_{pp}\geq0.5t_{dp}$ in the cuprates \cite{Hirayama18a,Hirayama19a}, where
$t_{dp}$ involves the $d_{x^2-y^2}$ orbitals. The $t_{dp}$ in state $|IV\rangle$, however, involves only the $d_{xy}$ orbitals and is hence  
considerably smaller than $t_{dp}$ in the cuprates. On the other hand, it is reasonable to assume that the magnitudes of $t_{pp}$ are similar in the
two classes of materials. Significantly enhanced conductivity can therefore occur from direct hopping between O-ions, but {\it only if charge carriers
occupy the O-sites to begin with.} Correspondingly,
a large contribution by $\Delta(W)$ to the stabilization of $|III\rangle$ with far larger number of charge carriers over $|IV\rangle$ with few
charge carriers on O-sites is to be anticipated.

(ii) The actual contribution by $\Delta(W)$ to the stabilization of $|III\rangle$ is even larger, given that in $|III\rangle$  $t_{dp}$ includes 
contributions from the $d_{x^2-y^2}$ orbitals, over and above from $d_{xy}$.

\begin{figure}
\includegraphics[width=3.0in]{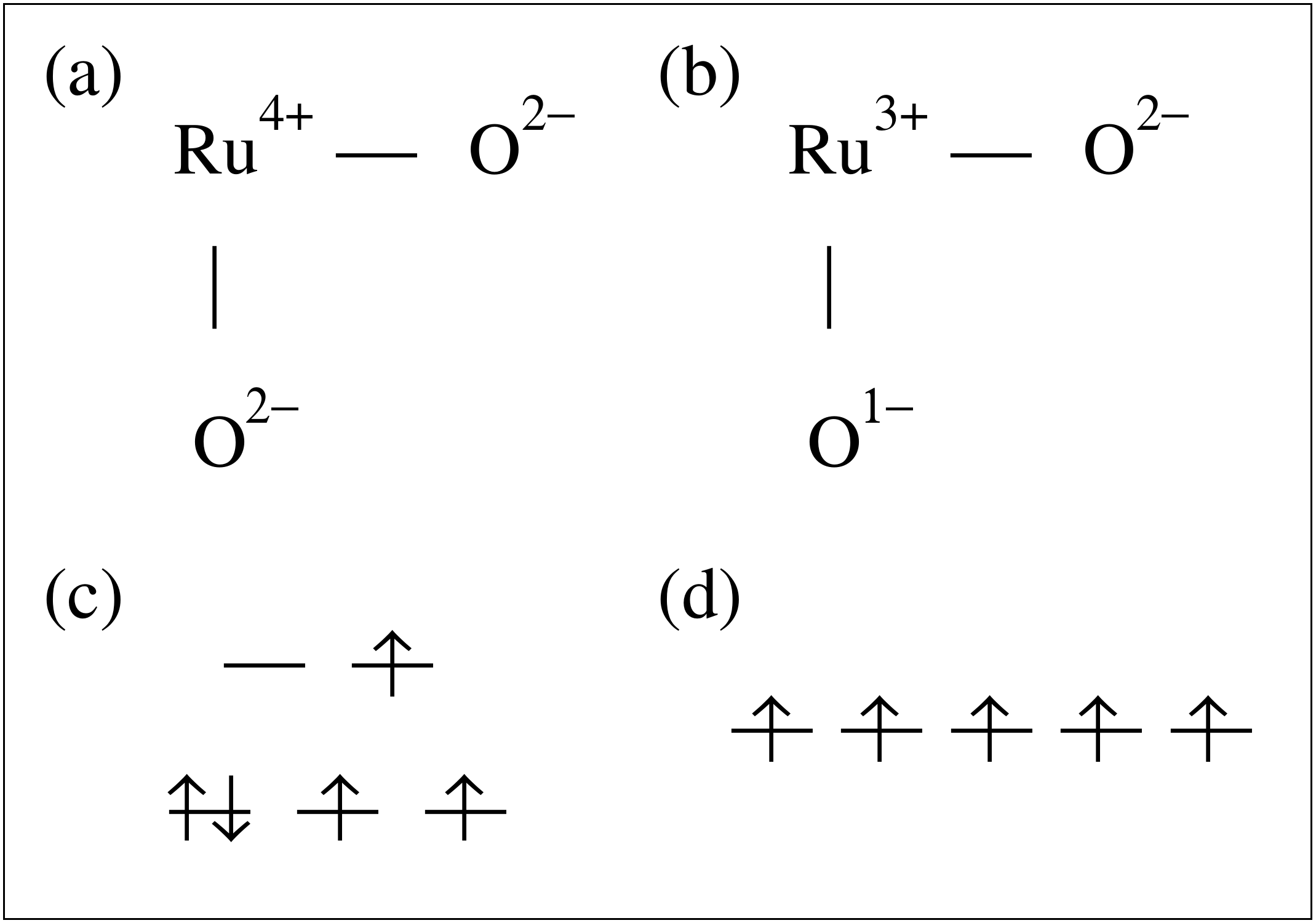}
\caption{Schematics of the layer intra-unit cell charge distributions in (a) insulating Ca$_2$RuO$_4$ and (b) metallic Sr$_2$RuO$_4$.
(c) Schematic of the virtual intermediate state in a current-induced (electron-doping induced) ruthenate conductor with an electron added to the low-spin Ru$^{4+}$ cation. For strong 
intra-orbital Coulomb repulsion the extra electron occupies an $e_g$ orbital and reduces the CFSE self-consistently, which in turn leads to (d) stable Ru$^{3+}$ ion 
in the metallic state with high ionization energy. Charge carriers in the metallic state are entirely on the O-ions in this case. Ru$^{3+}$ ions contribute to
magnetism but not transport.}
\label{IMT}
\end{figure}

(iii) $\Delta_{CF}$ in Eq.~2 need not be a rigid quantity as in the competition between two semiconductors, where $\Delta(W)=0$ by default.
When the competition is between a semiconductor and a metal it is likely that $\Delta_{CF}$ decreases self-consistently with conduction.
It is argued in (i) above that conduction due to electron hopping between O-ions requires state $|III\rangle$ to make significant contribution to the
electronic structure of Sr$_2$RuO$_4$. Given the large Coulomb repulsion between electrons occupying the same $d$-orbital, this implies that the
extra electron in the Sr$^{3+}$-ion will likely occupy an $e_g$ orbital (see Fig.~\ref{IMT}(c)), which in turn reduces $\Delta_{CF}$, leading to additional
occupancy of $e_g$ and finally to the configuration shown in Fig.~\ref{IMT}(d). 


Based on the above I posit that the difference between the Mott-Hubbard semiconductor bulk Ca$_2$RuO$_4$ and metallic Sr$_2$RuO$_4$ originates from 
different charges of the Ru-ions in these systems.
As a consequence of the negative charge-transfer gap in Sr$_2$RuO$_4$ there is a  preponderance of layer O$^{1-}$ ions instead of a few due to self-doping. 
In Figs.~\ref{IMT}(a) and (b) schematics of the IMT in oxides with large ionization energies of M$^{(n-1)+}$ are shown, while the schematics in Fig.~\ref{IMT}(c) iand (d) refer to the
current-induced Ru$^{4+} \to$ Ru$^{3+}$ transition. The difference between   
Ca$_2$RuO$_4$ and Sr$_2$RuO$_4$ is ascribed to the much smaller size of the Ca-ion leading to much larger Madelung energy stabilization of the high-charge state
and hence larger $\Delta E_{M,n}$ for Ca$_2$RuO$_4$ in Eq.~2, that favors
Ru$^{4+}$. It is shown in the next section that straightforward 
explanations of the experimental results presented in section II that are difficult to understand with low-spin Ru$^{4+}$
are obtained within the negative charge-transfer gap model in which Ru-ions occur as high-spin Ru$^{3+}$
which contribute to magnetism but not transport.

\section{\bf Explanations of Experiments within the negative charge-transfer gap model} 
\label{experiments}

\subsection{Sr$_2$RuO$_4$}
\label{SRO-explanation}

\begin{figure}
\includegraphics[width=3.0in]{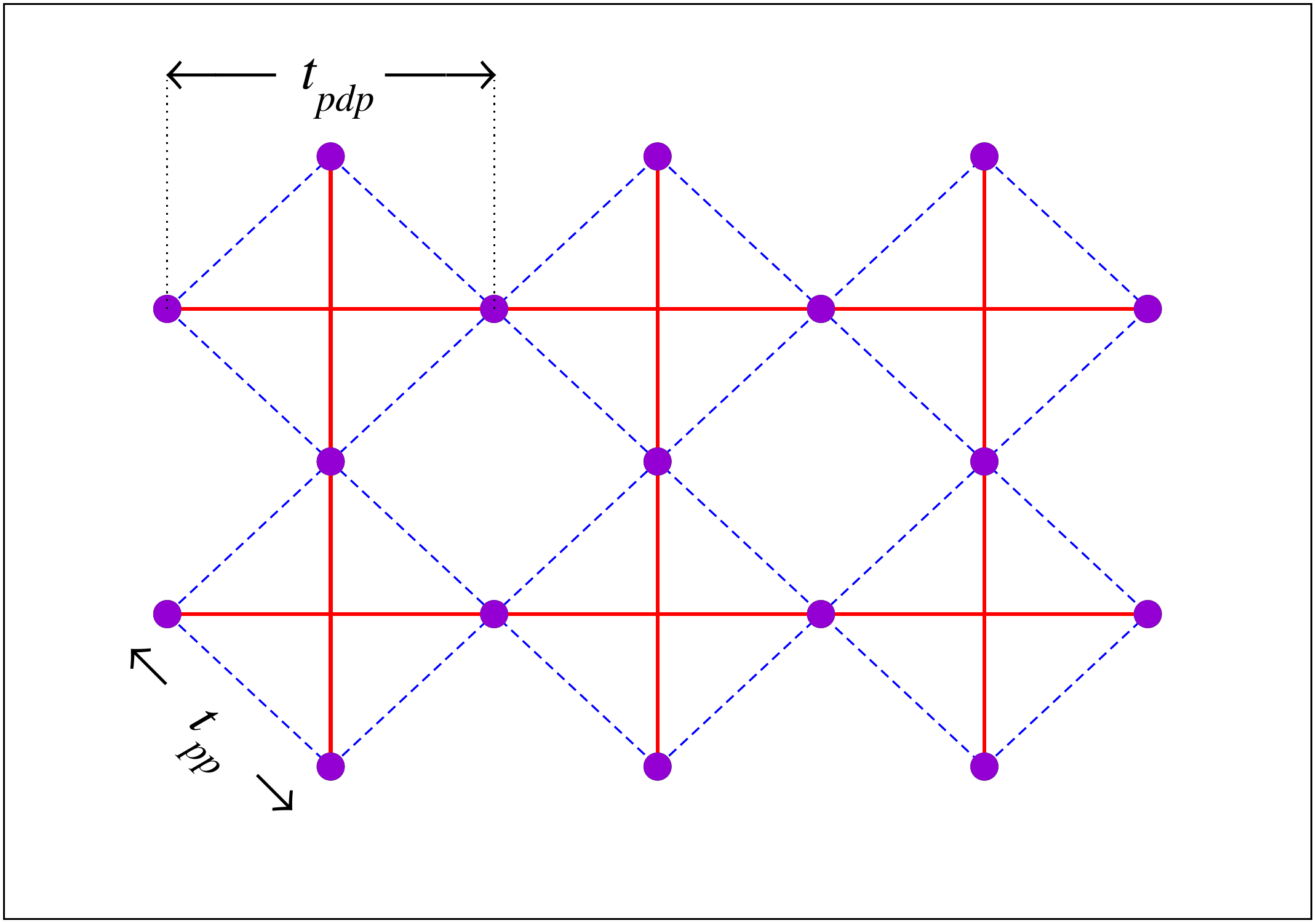}
\caption{The checkerboard lattice of O-ions, common to both Sr$_2$RuO$_4$ and cuprates, in their superconducting states.
The half-filled Ru$^{3+}$ (closed-shell Cu$^{1+}$) cations, which occur at the intersections of the solid lines,
are not explicitly shown as they do not play any role in superconductivity, which involves only the $\frac{3}{4}$-filled O-band.
The nearest neighbor direct O-O hoppings $t_{pp}$, as well as the O-O hoppings via the metal cations, $t_{pdp}$ (see text), are
shown. The magnitudes of $t_{pdp}$ are the same for O-M-O bond angles of 90$^o$ and 180$^o$, and are much larger for cuprates than for
Sr$_2$RuO$_4$. In the latter it is anticipated that $|t_{pp}|>|t_{pdp}|$ (see text).}
\label{checkerboard}
\end{figure}

\noindent {\it (i) $d$-wave SC.} Within the valence transition model the charge-carriers are entirely on the O-ions which have average charge $-1.5$. The Ru$^{3+}$-ions
play at most a virtual role in transport, exactly as the closed-shell Cu$^{1+}$ ions in the cuprates \cite{Mazumdar18a}. Fig.~\ref{checkerboard} shows 
the charge-carrying
checkerboard O-sublattice of the RuO$_2$ layer, rotated 45$^o$  relative to the Ru-O-Ru bonds. The
effective Hamiltonian $H_{eff}$ for the charge carriers on the checkerboard lattice is,
\begin{eqnarray}
 H_{eff} = -\sum_{\langle ij \rangle,\sigma}t_{pp} (p^\dagger_{i,\sigma}p_{j,\sigma}+H.c.) \\
\nonumber 
-\sum_{[ij],\sigma}t_{pdp} (p^\dagger_{i,\sigma}p_{j,\sigma}+H.c.) +U_p\sum_i n_{pi,\uparrow}n_{pi,\downarrow}\\
\nonumber
\frac{1}{2}\sum_{\langle i j\rangle} V_p^{NN} n_{pi} n_{pj} + \frac{1}{2}\sum_{(i j)} V_p^{NNN} n_{pi} n_{pj}\\
\nonumber
\label{hamuv}
\end{eqnarray}
Here $p^\dagger_{i,\sigma}$ creates a hole (O$^{1-}$) on the $p$-orbital of an O$^{2-}$ ion, $n_{pi,\sigma}=p^\dagger_{i,\sigma}p_{i,\sigma}$ and
$n_p=\sum_{\sigma=\uparrow,\downarrow}n_{pi,\sigma}$.
Sums are over the O-ions in the RuO$_2$ layer, $\langle~~\rangle$ denotes nearest neighbor (nn) oxygens;
$[~~]$ denotes O-ions linked via the same Ru$^{3+}$-ion (nn and nnn O-ions are linked by Ru-O bonds at
90$^o$ and 180$^o$ degrees, respectively); and $(~~)$ are nnn O-ions irrespective of whether they are linked via the same Ru-ion or not.
$U_p$, $V_p^{NN}$ and $V_p^{NNN}$ are Coulomb repulsions between pairs of holes on the same, nn and nnn $p$-orbitals,
respectively. The hopping parameter $t_{pp}$ is the direct hopping between nn O-ions while $t_{pdp}$ is the effective hopping between nn and nnn O-ions linked
by the same Ru$^{3+}$ ion, $t_{pdp}=t_{dp}^2/\Delta E$, $t_{dp}$ is the hopping between a Ru $d$-orbital and oxygen $p$-orbital and
$\Delta E$ = E(Ru$^{4+}$O$^{2-}$)$-$E(Ru$^{3+}$O$^{1-}$). The large ionization energy of Ru$^{3+}$ implies large $\Delta E$, which is likely to make         
$|t_{pdp}|<|t_{pp}|$. 

Given the $S=5/2$ spin state of Ru$^{3+}$, for $t_{pp}=0$ 
ferromagnetic spin coupling between O$^{1-}-$O$^{1-}$ would have been expected.
For $|t_{pp}|>|t_{pdp}|$ anticipated here, spin singlet coupling dominates. 
Sophisticated quantum mechanical calculations by the present author and colleagues \cite{Gomes16a,DeSilva16a,Clay19a} using 
exact diagonalization, Constrained Path Quantum Monte Carlo and Path Intergral 
Renormalization Group calculations have consistently shown that singlet superconducting pair-pair correlations are uniquely enhanced by the 
Hubbard interaction (relative to the noninteracting model) at $\frac{3}{4}$-filling and a narrow carrier density region about it on a 
geometrically frustrated lattice. 
At all other fillings the Hubbard interaction suppresses the pair-pair correlations 
relative to the noninteracting ($U_p=V_p^{NN}=V_p^{NNN}=0$) Hamiltonian, a result that agrees with the conclusions of reference \onlinecite{Qin19a}. 
Very recently, Clay and Roy have further shown that again uniquely at this same filling, Su-Schrieffer-Heeger bond phonons and the Hubbard $U$ act co-operatively
to further enhance the superconducting pair-pair correlation, while no such co-operative interaction is found at any other filling \cite{Clay20a}. 

Correlation-driven SC in the frustrated $\frac{3}{4}$-filled or $\frac{1}{4}$-filled band is {\it not} due to spin fluctuation. Rather, the 
superconducting state in the frustrated $\frac{3}{4}$-filled band evolves from a commensurate CDW of Cooper pairs, - a paired-electron crystal (PEC) -  
that is unique to the exactly $\frac{3}{4}$-filled frustrated lattice \cite{Li10a,Dayal11a}. Beyond cuprates, this theory of correlated-electron SC
readily explains the limitation of SC \cite{Plumb16a,Pei90a} to relatively narrow carrier concentration range $0.37 \leq x \leq 0.5$ in Ba$_{1-x}$K$_x$BiO$_3$, 
for which as of now there exists no other clear explanation. With the Bi-ions occurring only as Bi$^{3+}$, as determined experimentally 
\cite{Plumb16a,Khazraie18a}
the charge on the O-ions ranges from 1.50-1.54 for this range of $x$. The $s$-wave symmetry here is a natural consequence of the three-dimensional O-lattice
in  Ba$_{1-x}$K$_x$BiO$_3$. 

\noindent {\it (ii) Muon-spin rotation and apparent time-reversal breaking.} The apparent broken time reversal symmetry is associated with the
ferromagnetic Ru$^{3+}$-Ru$^{3+}$ spin-spin coupling, which, however, is unrelated to the SC itself. This is also the simplest explanation for the invariance
of the Josephson critical current under the inversion of current and magnetic fields \cite{Kashiwaya19a}: SC involves the O-ions only and not the Ru-ions. It is
very likely that the same mechanism of transport applies also to the ferromagnetic metal SrRuO$_3$, which has been thought to be an itinerant ferromagnet, 
A model-specific prediction is made
in the next section. 
  
{\it (iii) T$_c$ enhancement by the application if uniaxial pressure.} Carrier density-dependent calculations of superconducting pair-pair correlations 
have found that these are enhanced relative to the noninteracting Hamiltonian over a region of small width about
$\frac{3}{4}$-filling, with the strongest correlations occurring for exactly $\frac{3}{4}$-filling \cite{Gomes13a,DeSilva16a,Clay19a}. 
The implication for this is that should the filling be less than exactly
$\frac{3}{4}$ (as it should be under ambient pressure, since integral charges require complete reverse 
charge transfer Ru$^{4+}$(O$^{2-}$)$_2 \to$ Ru$^{3+}$(O$^{2-}$)$_1$(O$^{1-}$)$_1$)
T$_c$ will be less than the maximum possible. Within the valence transition model, pressure along the Ru-O bonds (but not along the Ru-Ru diagonal direction) 
enhances the reverse charge-transfer because of the increase in $t_{dp}$.   

\subsection{Ca$_2$RuO$_4$}

{\it (i) Current-induced IMT and diamagnetism.} There is no explanation of the current-induced IMT \cite{Nakamura13a,Okazaki13b} and diamagnetism \cite{Sow17a} 
within the usual models of Mott transition. 
Within the valence transition model the IMT is driven by the large $\Delta(W)$ within Eq.~\ref{1/2-filled}, giving 
a current-carrying state whose structure is very different from the insulator (see Fig.~\ref{IMT}), with nearly half the O-ions occurring as O$^{1-}$.
As mentioned above, the correlated $\frac{3}{4}$-filled geometrically frustrated lattice exhibits a strong tendency
to form a commensurate paired-electron crystal (PEC) \cite{Li10a,Dayal11a}, with nearest neighbor spin-singlets separated by pairs of vacancies (corresponding to 
periodic O$^{1-}$-O$^{1-}$-O$^{2-}$-O$^{2-}$). Reference \onlinecite{Mazumdar18a} has pointed out that the charge-ordered state
in the cuprates (and many other systems with the same carrier density) 
can be understood within the same picture. The concept of the PEC is identical to the concept of density-wave of Cooper pairs that has been conjectured by many authors to
be proximate to the superconducting state in the cuprates \cite{Franz04a,Chen04a,Tesanovic04a,Hamidian16a}. With increased frustration, the  PEC gives way to a paired-electron liquid and ultimately to a superconductor \cite{Gomes13a,DeSilva16a,Clay19a}.
The diamagnetism in the current-driven semimetallic state in Ca$_2$RuO$_4$ is ascribed to the paired-electron liquid, in agreement with prior 
explanation by the author of the low temperature Nernst effect measurements in cuprates \cite{Mazumdar18a}. 

{\it (ii) La-doping, IMT and ferromagnetism.} The valence transition model gives the simplest explanation for the IMT induced by La-doping, as illustrated
in Fig.~3(c) and (d). The extra electron initially occupies an $e_g$ orbital. Undoped Ca$_2$RuO$_4$ is already very close to instability,
and the smallest perturbation leads to a first-order transition. The ferromagnetism detected at small
dopings \cite{Cao00a} is due to the ferromagnetic couplings between the high-spin Ru-ions, which, as mentioned above, also explains the muon spin experiments
in Sr$_2$RuO$_4$.
 
{\it (iii) Co-appearance of ferromagnetism and possible SC.} This also has a natural explanation within the negative charge-transfer gap model. 
Observation of SC here is reminescent of superconductivity in undoped thin film T$^\prime$ cuprates, which is readily understood with the negative 
charge-transfer gap model \cite{Mazumdar18a}. Due to the reduced $\Delta E_{M,n}$ in thin films, the charge-transfer gap is naturally negative in these compounds.
Reduced $\Delta E_{M,n}$ is also true in nanocrystals of Ca$_2$RuO$_4$ and the $L$ as opposed to $S$ crystal structure is evidence for the same.   
Ferromagnetism driving SC, as claimed in Reference \onlinecite{Nobukane20a}, is a consequence of the requirement that SC can occur only when the
Ru-ions are in the trivalent state. The magnetism and the SC, however, involve two different components of the nanocrystal.

\section{Experimental Predictions.}

Here I make experimental predictions specific to the negative charge-transfer gap model.
\vskip 1pc
\noindent (i) {\it Charge densities on the layer oxygens in Sr$_2$RuO$_4$.} $^{17}$O NMR experiments should be able to find the charge densities on the layer oxygens. It is predicted that this
charge density in Sr$_2$RuO$_4$ in the superconducting state is $-1.5$.

\noindent (ii) {\it Intra-unit cell inequivalency of layer oxygens.} It is conceivable that in the current-induced semimetallic state, or in weakly La-doped CaRu$_2$O$_4$ intra-unit cell
inequivalency of layer O-ions will be detected by $^{17}$O-NMR or other experiments, as in the case of
YBa$_2$Cu$_3$O$_y$ \cite{Wu15a,Reichardt18a}.

\noindent (iii) {\it Spin susceptibility due to Ru-ions.} As of writing Knight-shift measurements have been repeated only for the O-sites \cite{Luo19a,Pustogow19a}. 
The earlier literature also reported extensive measurements of Ru-ion
spin susceptibility through T$_c$. $^{99}$Ru Knight-shift measurement with the magnetic field parallel to the RuO$_2$ layer \cite{Ishida01a}, 
and $^{101}$Ru Knight-shift measurements with the field perpendicular to the layer \cite{Ishida08a} both showed that the spin susceptibility due to Ru-ions remained unchanged
below T$_c$. {\it It is predicted that these observations will continue to be the same, in spite of the $^{17}$O NMR measurements.} 
If found to be true this will be the strongest proof that the Ru-ions do not participate
in normal state transport and SC in Sr$_2$RuO$_4$. 

\section {Conclusions}

In conclusion, nominal and true charges in conducting perovskite oxides can be very different. The natural state of oxide ions is not necessarily O$^{2-}$, even though
that is commonly assumed. The
second electron affinity of O is positive (it costs energy to add the second extra electron) and it is only the gain in Madelung energy in the insulating oxides that drives
a metal oxide to a state with high cation charge M$^{n+}$ and anion charge O$^{2-}$. With cations that in their lower charge 
state M$^{(n-1)+}$ have high ionization energies, IMT occurs via a
valence transition M$^{n+} \to$ M$^{(n-1)+}$, which in the quasi-2D materials leaves the layers with $\frac{3}{4}$-filled oxygen band of electrons. 
Average charge
of -1.5 on oxygen anions is neither strange nor unprecedented, - this is, for example, the known true charge on the oxygens in
alkali metal sequioxides \cite{Adler18a,Colman19a,Knaflic20a} X$_4$O$_6$, X = Rb, Cs. 

The negative charge-transfer gap model  gives the simplest yet most comprehensive explanations for the many apparently peculiar observations in Sr$_2$RuO$_4$ and Ca$_2$RuO$_4$. 
All of the peculiarities are understood once it is recognized that magnetism and SC involve two entirely different components of the crystals: the high-spin Ru$^{3+}$ ions
are responsible for the magnetic behavior but the charge carriers are entirely on the O-sites. The two phenomena are, however, coupled in that it is the lower charge
on the cation that gives the specific carrier density on the oxygen sites necessary for correlated-electron SC. Additonal attractive features of the theory are that
the model appears to be simultaneously applicable to many other correlated-electron systems \cite{Mazumdar18a} and is supported by numerical calculations of
superconducting pair-pair correlations \cite{Gomes16a,DeSilva16a,Clay19a,Clay20a}.

\section{Acknowledgments}

The author acknowledges support from NSF
grant CHE-1764152 and is grateful to Drs. R.
Torsten Clay (Mississippi State University), Charles Stafford (University of Arizona) and Shufeng Zhang (University of Arizona) 
for their careful readings of the manuscript and suggestions.
The author also acknowledges close interactions and collaborations through the years 
with Dr. Clay.

\begin{thebibliography}{10}

\bibitem{MacKenzie03a}
A.~P. Mackenzie and Y.~Maeno.
\newblock The superconductivity of \protect{Sr$_2$RuO$_4$} and the physics of
  spin-triplet pairing.
\newblock {\em Rev.\ Mod.\ Phys.}, 75:657--712, 2003.

\bibitem{Mackenzie17a}
A.~P. Mackenzie, T.~Scaffidi, C.~W. Hicks, and Y.~Maeno.
\newblock Even odder after twnety-three years: the superconducting order
  parameters puzzle of \protect{Sr$_2$RuO$_4$}.
\newblock {\em npj Quantum Materials}, 2017.

\bibitem{Hicks14a}
C.~W. Hicks, D.~O. Brodsky, E.~A. Yelland, A.~S. Gibbs, J.~A.~N. Bruin, M.~E.
  Barber, S.~D. Edkins, K.~Nishimura, S.~Yonezawa, Y.~Maeno, and A.~P.
  Mackenzie.
\newblock Strong increase of \protect{$T_c$} of \protect{Sr$_2$RuO$_4$} under
  both tensile and compressive strain.
\newblock {\em Science}, 344:283--285, 2014.

\bibitem{Taniguchi15a}
H.~Taniguchi, K.~Nishimura, S.~K. Goh, S.~Yonezawa, and Y.~Maeno.
\newblock Higher-\protect{$T_c$} superconducting phase in
  \protect{Sr$_2$RuO$_4$} induced by in-plane uniaxial pressure.
\newblock {\em J.\ Phys.\ Soc.\ Jpn.}, 84:014707, 2015.

\bibitem{Steppke17a}
A.~Steppke, L.~Zhao, M.~E. Barber, T.~Scaffidi, F.~Jerzembeck, H.~Rosner, A.~S.
  Gibbs, Y.~Maeno, S.~H. Simon, A.~P. Mackenzie, and C.~W. Hicks.
\newblock Strong peak in \protect{$T_c$} of \protect{Sr$_2$RuO$_4$} under
  uniaxial pressure.
\newblock {\em Science}, 355:eaaf9398, 2017.

\bibitem{Hassinger17a}
E.~Hassinger, P.~Bourgeois-Hope, H.~Taniguchi, S.~R. de~Cotret,
  G.~Grissonnanche, M.~S. Anwar, Y.~Maeno, N.~Doiron-Leyraud, and L.~Taillefer.
\newblock Vertical line nodes in the superconducting gap structure of
  \protect{Sr$_2$RuO$_4$}.
\newblock {\em Phys. Rev. X}, 7:011032, 2017.

\bibitem{Luo19a}
Y.~Luo et~al.
\newblock Normal state \protect{$^{17}$O} nmr studies of
  \protect{Sr$_2$RuO$_4$} under uniaxial stress.
\newblock {\em Phys. Rev. X}, 9:021044, 2019.

\bibitem{Pustogow19a}
A.~Pustogow, Yongkang Luo, A.~Chronister, Y.-S. Su, D.~A. Sokolov,
  F.~Jerzembeck, A.~P. Mackenzie, C.~W. Hicks, N.~Kikugawa, S.~Raghu, E.~D.
  Bauer, and S.~E. Brown.
\newblock Constraints on the superconducting order parameter in
  \protect{Sr$_2$RuO$_4$} from oxygen-17 nuclear magnetic resonance.
\newblock {\em Nature}, 574:72--75, 2019.

\bibitem{Sharma20a}
R.~Sharma, S.~D. Edkins, Z.~Wang, A.~Kostina, C.~Sow, Y.~Maeno, A.~P.
  Mackenzie, J.~C.~Seamus Davis, and V.~Madhavan.
\newblock Momentum-resolved superconducting energy gaps of
  \protect{Sr$_2$RuO$_4$} from quasiparticle interference imaging.
\newblock {\em Proc. Natl. Acad. Sci. USA}, 117:5222--5227, 2020.

\bibitem{Mazumdar18a}
S.~Mazumdar.
\newblock Valence transition model of the pseudogap, charge order, and
  superconductivity in electron-doped and hole-doped copper oxides.
\newblock {\em Phys.\ Rev.\ B}, 98:205153, 2018.

\bibitem{Torre15a}
A.~de~la Torre, S.~McKeown Walker, F.~Y. Bruno, S.~Ricco, Z.~Wang, I.~Gutierrez
  Lezama, G.~Scheerer, G.~Giriat, D.~Jacard, C.~Berthod, T.~K. Kim, M.~Hoesch,
  E.~C. Hunter, R.~S. Perry, A.~Tamai, and F.~Baumberger.
\newblock Collapse of the \protect{Mott} gap and emergence of a nodal liquid in
  lightly doped \protect{Sr$_2$RuO$_4$}.
\newblock {\em Phys.\ Rev.\ Lett.}, 115:176402, 2015.

\bibitem{Kim16a}
Y.~H. Kim, N.~H. Sung, J.~D. Denlinger, and B.~J. Kim.
\newblock Observation of a \protect{$d$}-wave gap in electron-doped
  \protect{Sr$_2$IrO$_4$}.
\newblock {\em Nat.\ Phys.}, 12:37--42, 2016.

\bibitem{Battisti17a}
I.~Battisti, K.~M. Bastiaans, V.~Fedoseev, A.~de~la Torre, N.~Iliopoulos,
  A.~Tamai, E.~C. Hunter, R.~S. Perry, J.~Zaanen, F.~Baumberger, and M.~P.
  Allan.
\newblock Universality of pseudogap and emergent order in lightly doped
  \protect{M}ott insulators.
\newblock {\em Nat.\ Phys.}, 13:21--26, 2017.

\bibitem{Torrance81a}
J.~B. Torrance, J.~E. Vazquez, J.~J. Mayerle, and V.~Y. Lee.
\newblock Discovery of a neutral-to-ionic phase-transition in organic
  materials.
\newblock {\em Phys.\ Rev.\ Lett.}, 46:253--257, 1981.

\bibitem{Kosihara90a}
S.~Kosihara, Y.~Tokura, T.~Mitani, G.~Saito, and T.~Koda.
\newblock Photoinduced valence instability in the organic molecular compound
  tetrathiafulvalene-p-chloranil.
\newblock {\em Phys.\ Rev.\ B}, 42:6853--6856, 1990.

\bibitem{Masino17a}
M.~Masino, N.~Castagnetti, and A.~Girlando.
\newblock Phenomenology of the neutral-ionic valence instability in mixed stack
  charge-transfer crystals.
\newblock {\em Crystals}, 7:108, 2017.

\bibitem{Scheerer18a}
G.~W. Scheerer, Z.~Ren, S.~Watanabe, G.~Lapertot, D.~Aoki, D.~Jaccard, and
  K.~Miyake.
\newblock The dominant role of critical valence fluctuations on high t$_c$
  superconductivity in heavy fermions.
\newblock {\em NPJ Quant. Mater.}, 3:41, 2018.
\bibitem{Nakamura13a}
F.~Nakamura, M.~Sakaki, Y.~Yamanaka, S.~Tamaru, T.~Suzuki, and Y.~Maeno.
\newblock Electric-field-induced metal maintained by current of the mott
  insulator \protect{Ca$_2$RuO$_4$}.
\newblock {\em Sci. Rep.}, 3:2536, 2013.

\bibitem{Okazaki13b}
R.~Okazaki, Y.~Nishina, Y.~Yasui, F.~Nakamura, T.~Suzuki, and I.~Terasaki.
\newblock Current-induced gap suppression in the the mott insulator
  \protect{Ca$_2$RuO$_4$}.
\newblock {\em J.\ Phys.\ Soc.\ Jpn.}, 82:103702, 2013.

\bibitem{Cao00a}
C.~Cao, S.~McCall, V.~Dobrosavljevic, C.~S. Alexander, J.~E. Crow, and R.~P.
  Guertin.
\newblock Grond-state instability of the mott insulator
  \protect{Ca$_2$RuO$_4$}: Impact of light la doping on the metal-insulator
  transition and magnetic ordering.
\newblock {\em Phys.\ Rev.\ B}, 61:R5053--R5057, 2000.

\bibitem{Nobukane20a}
H.~Nobukane, K.~Yanagihara, Y.~Kunisada, Y.~Ogasawara, K.~Isono, K.~Nomura,
  K.~Tanahashi, T.~Nomura, T.~Akiyama, and S.~Tanda.
\newblock Co-appearance of superconductivity and ferromagnetism in a
  \protect{Ca$_2$RuO$_4$} nanofilm crystal.
\newblock {\em Sci. Rep.}, 10:3462, 2020.

\bibitem{Ishida98a}
K.~Ishida, H.~Mukuda, Y.~Kitaoka, K.~Asayama, Z.~Q. Mao, Y.~Mori, and Y.~Maeno.
\newblock Spin-triplet superconductivity in \protect{Sr$_2$RuO$_4$} identified
  by \protect{$^{17}$O} knight shift.
\newblock {\em Nature}, 396:658--660, 1998.

\bibitem{Luke98a}
G.~M. Luke, Y.~Fudamoto, K.~M. Kojima, M.~I. Larkin, J.~Merrin, B.~Nachumi,
  Y.~J. Uemura, Y.~Maeno, Z.~Q. Mao, Y.~Mori, H.~Nakamura, and M.~Sigrist.
\newblock Time-reversal symmetry-breaking superconductivity in
  \protect{Sr$_2$RuO$_4$}.
\newblock {\em Nature}, 394:558--561, 1998.

\bibitem{Xia06a}
J.~Xia, Y.~Maeno, P.~T. Beyersdorf, M.~M. Fejer, and A.~Kapitulnik.
\newblock High resolution polar kerr effect measurements of
  \protect{Sr$_2$RuO$_4$}: Evidence for broken time-reversal symmetry in the
  superconducting state.
\newblock {\em Phys.\ Rev.\ Lett.}, 97:167002, 2006.

\bibitem{Kashiwaya19a}
S.~Kashiwaya, K.~Saitoh, H.~Kashiwaya~M. Koyanagi, M.~Sato, K.~Yada, and
  Y.~Tanakaand~Y. Maeno.
\newblock Time-reversal invariant superconductivity of \protect{Sr$_2$RuO$_4$}
  revealed by josephson effects.
\newblock {\em Phys.\ Rev.\ B}, 100:094530, 2019.

\bibitem{Karapetyan12a}
H.~Karapetyan, M.~H{\"u}cker, G.~D. Gu, J.~M. Tranquada, M.~M. Fejer, J.~Xia,
  and A.~Kapitulnik.
\newblock Magneto-optical measurements of a cascade of transitions in
  superconducting \protect{La$_{1:875}$Ba$_{0:125}$CuO$_4$} single crystals.
\newblock {\em Phys.\ Rev.\ Lett.}, 109:147001, 2012.

\bibitem{Grinenko20a}
V.~Grinenko et~al.
\newblock Split superconducting and time-reversal symmetry-breaking
  transitions, and magnetic order in \protect{Sr$_2$RuO$_4$} under uniaxial
  stress.
\newblock {\em arXiv:2001.08152v2}, 2020.

\bibitem{Yonezawa13a}
S.~Yonezawa, T.~Kajikawa, and Y.~Maeno.
\newblock First-order superconducting transition of \protect{Sr$_2$RuO$_4$}.
\newblock {\em Phys.\ Rev.\ Lett.}, 110:077003, 2013.

\bibitem{Braden98a}
M.~Braden, G.~And\r'e, S.~Nakatsuji, and Y.~Maeno.
\newblock Crystal and magnetic structure of \protect{Ca$_2$RuO$_4$}:
  Magnetoelastic coupling and the metal-insulator transition.
\newblock {\em Phys.\ Rev.\ B}, 58:847--861, 1998.

\bibitem{Friedt01a}
O.~Friedt, M.~Braden, G.~And\'re, P.~Adelmann, S.~Nakatsuji, and Y.~Maeno.
\newblock Structural and magnetic aspects of the metal-insulator transition in
  \protect{Ca$_{2-x}$Sr$_x$RuO$_4$}.
\newblock {\em Phys.\ Rev.\ B}, 63:174432, 2001.

\bibitem{Fukazawa01a}
H.~Fukazawa1 and Y.~Maeno.
\newblock Filling control of the mott insulator \protect{Ca$_2$RuO$_4$}.
\newblock {\em J.\ Phys.\ Soc.\ Jpn.}, 70:460--467, 2001.

\bibitem{Nakatsuji04a}
S.~Nakatsuji, V.~Dobrosavljevi\'c, D.~Tanaskovi\'c, M.~Minakata, H.~Fukazawa,
  and Y.~Maeno.
\newblock Mechanism of hopping transport in disordered mott insulators.
\newblock {\em Phys.\ Rev.\ Lett.}, 93:146401, 2004.

\bibitem{Liebsch07a}
A.~Liebsch and H.~Ishida.
\newblock Subband filling and mott transiton in
  \protect{Ca$_{2-x}$Sr$_x$RuO$_4$}.
\newblock {\em Phys.\ Rev.\ Lett.}, 98:216403, 2007.

\bibitem{Gorelov10a}
E.~Gorelov, M.~Karolak, T.~O. Wehling, F.~Lechermann, A.~I. Lichtenstein, and
  E.~Pavarini.
\newblock Nature of the mott transition in \protect{Ca$_2$RuO$_4$}.
\newblock {\em Phys.\ Rev.\ Lett.}, 104:226401, 2010.

\bibitem{Nakamura02b}
F.~Nakamura, T.~Goko, M.~Ito, T.~Fujita, S.~Nakatsuji, H.~Fukazawa, Y.~Maeno,
  P.~Alireza, D.~Forsythe, and S.~R. Julian.
\newblock From mott insulator to ferromagnetic metal: A pressure study of
  \protect{Ca$_2$RuO$_4$}.
\newblock {\em Phys.\ Rev.\ B}, 65:220402(R), 2002.
\bibitem{Sow17a}
C.~Sow, S.~Yonezawa, S.~Kitamura, T.~Oka, K.~Kuroki, F.~Nakamura, and Y.~Maeno.
\newblock Current-induced strong diamagnetism in the mott insulator
  \protect{Ca$_2$RuO$_4$}.
\newblock {\em Science}, 358:1084--1087, 2017.

\bibitem{Fursich19a}
K.~F{\'u}rsich, J.~Bertinshaw, P.~Butler, M.~Krautloher, M.~Minola, and
  B.~Keimer.
\newblock Raman scattering from current-stabilized nonequlibrium phases in
  \protect{Ca$_2$RuO$_4$}.
\newblock {\em Phys.\ Rev.\ B}, 100:081101(R), 2019.

\bibitem{Zhang19a}
J.~Zhang et~al.
\newblock Nano-resolved current-induced insulator-metal transition in the mott
  insulator \protect{Ca$_2$RuO$_4$}.
\newblock {\em Phys. Rev. X}, 9:011032, 2019.

\bibitem{Bertinshaw19a}
J.~Bertinshaw et~al.
\newblock Unique crystal structure of \protect{Ca$_2$RuO$_4$} in the current
  stabilized semimetallic state.
\newblock {\em Phys.\ Rev.\ Lett.}, 123:137204, 2019.

\bibitem{Zhao19a}
H.~Zhao, B.~Hu, F.~Ye, C.~Hoffmann, I.~Kimchi, and G.~Cao.
\newblock Nonequlibrium orbital transitions via applied electrical current in
  calcium ruthenates.
\newblock {\em Phys.\ Rev.\ B}, 100:241104(R), 2019.

\bibitem{Naito16a}
M.~Naito, Y.~Krockenberger, A.~Ikeda, and H.~Yamamoto.
\newblock Reassessment of the electronic state, magnetism, and
  superconductivity in high-\protect{T$_c$} cuprates with the
  \protect{Nd$_2$CuO$_4$} structure.
\newblock {\em Physica C}, 523:28--54, 2016.

\bibitem{Wu15a}
T.~Wu, H.~Mayaffre, S.~Kraemer, M.~\protect{Horvati\'c}, C.~Berthier, W.~N.
  Hardy, L.~Ruixing, D.~A. Bonn, and M.-H. Julien.
\newblock Incipient charge order observed by \protect{NMR} in the normal state
  of \protect{YBa$_2$Cu$_3$O$_y$}.
\newblock {\em Nat.\ Commun.}, 6:6438, 2015.

\bibitem{Kohsaka12a}
Y.~Kohsaka, T.~Hanaguri, M.~Azuma, M.~Takano, J.~C. Davis, and H.~Takagi.
\newblock Visualization of the emergence of the pseudogap state and the
  evolution to superconductivity in a lightly hole-doped mott insulator.
\newblock {\em Nat. Phys.}, 8:534--538, 2012.

\bibitem{Comin15a}
R.~Comin, R.~Sutarto, F.~He, E.~H. D.~S. Neto, L.~Chauviere, A.~Frano,
  R.~Liang, W.~N. Hardy, D.~A. Bonn, Y.~Yoshida, H.~Eisaki, A.~J. Achkar, D.~G.
  Hawthorn, B.~Keimer, G.~A. Sawatzky, and A.~Damascelli.
\newblock Symmetry of charge order in cuprates.
\newblock {\em Nature Materials}, 14:796--800, 2015.

\bibitem{Reichardt18a}
S.~Reichardt, M.~Jurkutat, R.~Guehne, J.~Kohlrautz, A.~Erb, and J.~Haase.
\newblock Proof of bulk charge ordering in the \protect{CuO$_2$} plane of the
  cuprate superconductor \protect{YBa$_2$Cu$_3$O$_{6.9}$} by high pressure nmr.
\newblock {\em Condensed Matter}, 3:23, 2018.

\bibitem{Armitage10a}
N.~P. Armitage, P.~Fournier, and R.~L. Greene.
\newblock Progress and perspectives on electron-doped cuprates.
\newblock {\em Rev.\ Mod.\ Phys.}, 82:2421--2487, 2010.

\bibitem{SilvaNeto15a}
E.~H. D.~S. Neto, R.~Comin, F.~He, R.~Sutarto, Y.~Jiang, R.~L. Greene, G.~A.
  Sawatzky, and A.~Damascelli.
\newblock Charge ordering in the electron-doped {Nd$_{2-x}$Ce$_x$CuO$_4$}.
\newblock {\em Science}, 347:282--285, 2015.

\bibitem{SilvaNeto16a}
E.~H. D.~S. Neto, B.~Yu, M.~Minola, R.~Sutarto, E.~Schierle, F.~Boschini,
  M.~Zonno, M.~Bluschke, J.~Higgins, Y.~Li, G.~Yu, E.~Weschke, F.~He, M.~Le
  Tacon, R.~L. Greene, M.~Greven, G.~A. Sawatzky, B.~Keimer, and A.~Damascelli.
\newblock Doping-dependent charge order correlations in electron-doped
  cuprates.
\newblock {\em Sci. Adv}, 2:e1600782, 2016.

\bibitem{Rice81a}
T.~M. Rice and L.~Sneddon.
\newblock Real space electron pairing in \protect{BaPb$_{1-x}$Bi$_x$O$_3$}.
\newblock {\em Phys.\ Rev.\ Lett.}, 47:689--692, 1981.

\bibitem{Plumb16a}
N.~C. Plumb, D.~J. Gawryluk, Y.~Wang, Z.~Ristic, J.~Park, B.~Q. Lv, Z.~Wang,
  C.~E. Matt, N.~Xu, T.~Shang, K.~Conder, J.~Mesot, S.~Johnston, M.~Shi, and
  M.~Radovic.
\newblock Momentum-resolved electronic structure of the high-\protect{T$_c$}
  superconductor parent compound \protect{BaBiO$_3$}.
\newblock {\em Phys.\ Rev.\ Lett.}, 117:037002, 2016.

\bibitem{Khazraie18a}
A.~Khazraie, K.~Foyevtsova, I.~Elfimov, and G.~A. Sawatzky.
\newblock Oxygen holes and hybridization in the bismuthates.
\newblock {\em Phys.\ Rev.\ B}, 97:075103, 2018.

\bibitem{Pei90a}
S.~Pei, J.~D. Jorgensen, B.~Dabrowski, D.~G. Hinks, D.~R. Richards, and A.~W.
  Mitchell.
\newblock Structural phase diagram of the \protect{Ba$_{1-x}$K$_x$BiO$_3$}
  system.
\newblock {\em Phys.\ Rev.\ B}, 41:4126--4141, 1990.

\bibitem{Kim08a}
B.~J. Kim, H.~Jin, S.~J. Moon, J.-Y. Kim, B.-G. Park, C.~S. Leem, J.~Yu, T.~W.
  Noh, C.~Kim, S.-J. Oh, J.-H. Park, V.~Durairaj, G.~Cao, and E.~Rotenberg.
\newblock Novel \protect{J$_{\rm eff}$=1/2} {M}ott state induced by
  relativistic spin-orbit coupling in \protect{Sr$_2$IrO$_4$}.
\newblock {\em Phys.\ Rev.\ Lett.}, 101:076402, 2008.
\bibitem{Yan15a}
Y.~J. Yan, M.~Q. Ren, H.~C. Xu, B.~P. Xie, R.~Tao, H.~Y. Choi, N.~Lee, Y.~J.
  Choi, T.~Zhang, and D.~L. Feng.
\newblock Electron-doped \protect{Sr$_2$IrO$_4$}: An analogue of hole-doped
  cuprate superconductors demonstrated by scanning tunneling microscopy.
\newblock {\em Phys. Rev. X}, 5:041018, 2015.

\bibitem{Hashimoto14a}
M.~Hashimoto, I.~M. Vishik, R.-H. He, T.~P. Devereaux, and Z.-X. Shen.
\newblock Energy gaps in the high-transition temperature cuprate
  superconductors.
\newblock {\em Nat. Phys.}, 10:483--495, 2014.

\bibitem{Qin19a}
C.-M.~Chung M.~Qin, H.~Shi, E.~Vitali, C.~Hubig, S.~R.~White
  U.~\protect{Scholl\"ock}, and S.~Zhang.
\newblock Absence of superconductivity in the pure two-dimensional hubbard
  model.
\newblock {\em arXiv:1910.08931}, 2019.

\bibitem{Gomes16a}
N.~Gomes, W.~Wasanthi~De Silva, T.~Dutta, R.~T. Clay, and S.~Mazumdar.
\newblock Coulomb enhanced superconducting pair correlations in the frustrated
  quarter-filled band.
\newblock {\em Phys.\ Rev.\ B}, 93:165110, 2016.

\bibitem{Mazumdar78a}
S.~Mazumdar and Z.~G. Soos.
\newblock Neutral-ionic interface in organic charge-transfer salts.
\newblock {\em Phys.\ Rev.\ B}, 18:1991--2003, 1978.

\bibitem{Nagaosa86a}
N.~Nagaosa and J.~Takimoto.
\newblock Theory of neutral-ionic transition in organic crystals.1.
  \protect{Monte-Carlo} simulation of modified \protect{Hubbard} model.
\newblock {\em J.\ Phys.\ Soc.\ Jpn.}, 55:2735--2744, 1986.

\bibitem{Pati01a}
Y.~Anusooya-Pati, Z.~G. Soos, and A.~Painelli.
\newblock Symmetry crossover and excitation thresholds at the neutral-ionic
  transition of the modified \protect{H}ubbard model.
\newblock {\em Phys.\ Rev.\ B}, 63:205118, 2001.

\bibitem{DeSilva16a}
W.~Wasanthi~De Silva, N.~Gomes, S.~Mazumdar, and R.~T. Clay.
\newblock Coulomb enhancement of superconducting pair-pair correlations in a
  3/4-filled model for \protect{$\kappa$-(BEDT-TTF)$_{2}$X}.
\newblock {\em Phys.\ Rev.\ B}, 93:205111, 2016.

\bibitem{Clay19a}
R.~T. Clay and S.~Mazumdar.
\newblock From charge- and spin-ordering to superconductivity in the organic
  charge-transfer solids.
\newblock {\em Phys. Reports}, 788:1--89, 2019.

\bibitem{Oh13a}
Y.~S. Oh, J.~J. Yang, Y.~Horibe, and S.-W. Cheong.
\newblock Anionic depolymerization transition in \protect{IrTe$_2$}.
\newblock {\em Phys.\ Rev.\ Lett.}, 110:127209, 2013.

\bibitem{Fang13a}
A.~F. Fang, G.~Xu, T.~Dong, P.~Zheng, and N.~L. Wang.
\newblock Structural phase transition in \protect{IrTe$_2$}: a combined study
  of optical spectroscopy and band structure calculations.
\newblock {\em Sci. Rep.}, 3:1153, 2013.

\bibitem{Itoh95a}
M.~Itoh, M.~Shikano, and T.~Shimura.
\newblock High- and low-spin transition of \protect{Ru$^{4+}$} in the
  perovskite-related layered system \protect{Sr$_{n+1}$Ru$_n$O$_{3n+1}$, $n=1$,
  2, and $\infty$} with changes in \protect{$n$}.
\newblock {\em Phys.\ Rev.\ B}, 51:16432--16435, 1995.

\bibitem{Grutter12a}
A.~J. Grutter, F.~J. Wong, E.~Arenholz, A.~Vailionis, and Y.~Suzuki.
\newblock Evidence of high-spin ru and universal magnetic anisotropy in
  \protect{SrRuO$_3$} thin films.
\newblock {\em Phys.\ Rev.\ B}, 85:134429, 2012.

\bibitem{Hirayama18a}
M.~Hirayama, Y.~Yamaji, T.~Misawa, and M.~Imada.
\newblock Ab inition effective hamiltonians for cuprate superconductors.
\newblock {\em Phys.\ Rev.\ B}, 98:134501, 2018.

\bibitem{Hirayama19a}
M.~Hirayama, T.~Misawa, T.~Ohgoe, Y.~Yamaji, and M.~Imada.
\newblock Effective hamiltonian for cuprate superconductors derived from
  multiscale ab initio scheme with level renormalization.
\newblock {\em Phys.\ Rev.\ B}, 99:245155, 2019.

\bibitem{Clay20a}
R.~T. Clay and D.~Roy.
\newblock Cooperative enhancement of superconducting correlations by
  electron-electron and electron-phonon interactions in the quarter-filled
  band.
\newblock {\em arXiv:1911.06158}, 2020.

\bibitem{Li10a}
H.~Li, R.~T. Clay, and S.~Mazumdar.
\newblock The paired-electron crystal in the two-dimensional frustrated
  quarter-filled band.
\newblock {\em J. Phys.: Condens. Matter}, 22:272201, 2010.

\bibitem{Dayal11a}
S.~Dayal, R.~T. Clay, H.~Li, and S.~Mazumdar.
\newblock Paired electron crystal: Order from frustration in the quarter-filled
  band.
\newblock {\em Phys.\ Rev.\ B}, 83:245106, 2011.

\bibitem{Gomes13a}
N.~Gomes, R.~T. Clay, and S.~Mazumdar.
\newblock Absence of superconductivity and valence bond order in the
  {H}ubbard-{H}eisenberg model for organic charge-transfer solids.
\newblock {\em J. Phys. Condens. Matter}, 25:385603, 2013.
\bibitem{Franz04a}
M.~Franz.
\newblock Crystalline electron pairs.
\newblock {\em Nature}, 305:1410--1411, 2004.

\bibitem{Chen04a}
Han-Dong Chen, Oskar Vafek, Ali Yazdani, and Shou-Cheng Zhang.
\newblock Pair density wave in the pseudogap state of high temperature
  superconductors.
\newblock {\em Phys.\ Rev.\ Lett.}, 93:187002, 2004.

\bibitem{Tesanovic04a}
Z.~Tesanovic.
\newblock Charge modulation, spin response, and dual {H}ofstadter butterfly in
  \protect{High-T$_c$} cuprates.
\newblock {\em Phys.\ Rev.\ Lett.}, 93:217004, 2004.

\bibitem{Hamidian16a}
M.~H. Hamidian, S.~D. Edkins, S.~H. Joo, A.~Kostin, H.~Eisaki, S.~Uchida, M.~J.
  Lawler, E.-A. Kim, A.~P. Mackenzie, K.~Fujita, J.~Lee, and J.~C. Davis.
\newblock Detection of a {C}ooper-pair density wave in
  \protect{Bi$_2$Sr$_2$CaCu$_2$O$_{8+x}$}.
\newblock {\em Nature}, 532:343--347, 2016.

\bibitem{Ishida01a}
K.~Ishida, H.~Mukuda, Y.~Kitaoka, Z.~Q. Mao, H.~Fukazawa, and Y.~Maeno.
\newblock Ru nmr probe of spin susceptibility in the superconducting state of
  \protect{Sr$_2$RuO$_4$}.
\newblock {\em Phys.\ Rev.\ B}, 63:060507(R), 2001.

\bibitem{Ishida08a}
K.~Ishida, H.~Murakawa, H.~Mukuda, Y.~Kitaoka, Z.~Q. Mao, and Y~Maeno.
\newblock Nmr and nqr studies on superconducting \protect{Sr$_2$RuO$_4$}.
\newblock {\em J. Phys. Chem. Solids}, 69:3108--3114, 2008.

\bibitem{Adler18a}
P.~Adler et~al.
\newblock Verwey-type charge ordering transition in an open-shell p-electron
  compound.
\newblock {\em Sci. Adv.}, 4:eaap7581, 2018.

\bibitem{Colman19a}
R.~H. Colman, H.~E. Okur, W.~Kockelmann, C.~M. Brown, S.~Annette, C.~Felser,
  M.~Jansen, and K.~Prassides.
\newblock Elusive valence transition in mixed-valence sesquioxide
  \protect{Cs$_4$O$_6$"}.
\newblock {\em Inorg. Chem.}, 58:14532--14541, 2019.

\bibitem{Knaflic20a}
T.~Knaflic, P.~Jeglic, M.~Komelj, A.~Zorko, P.~K. Biswas, A.~N. Ponomaryov,
  S.~A. Zvyagin, M.~Reehuis, A.~Hoser, M.~Geiss, J.~Janek, P.~Adler, C.~Felser,
  M.~Jansen, and nd~D.~Arcon.
\newblock Spin-dimer ground state driven by consecutive charge and orbital
  ordering transitions in the anionic mixed-valence compound
  \protect{Rb$_4$O$_6$}.
\newblock {\em Phys.\ Rev.\ B}, 101:024419, 2020.

\end{thebibliography}

\end{document}